\documentclass[journal,comsoc]{IEEEtran}

\usepackage{graphicx}
\usepackage{epstopdf}
\usepackage{amsmath}
\usepackage{times}
\usepackage{latexsym}
\usepackage{bm}

\usepackage{cite}
\usepackage[numbers,sort&compress]{natbib}

\usepackage{amssymb}
\usepackage[center]{caption2}
\usepackage{stfloats}
\usepackage{cases}
\usepackage{subfigure}

\usepackage{array}
\usepackage{setspace}
\usepackage{fancyhdr}
\usepackage{colortbl}
\usepackage{CJK}
\usepackage{tikz}
\usepackage[noend]{algpseudocode}

\usepackage{algorithmicx,algorithm}
\newtheorem{remark}{Remark}

\allowdisplaybreaks[4]
\author{Chenlan~Lin,~Xiaoming~Chen,~and~Zhaoyang~Zhang% <-this % stops a space
\thanks{C.~Lin,~X.~Chen,~and~Z.~Zhang are with the College of Information Science and Electronic Engineering, Zhejiang University, Hangzhou 310027, China (e-mails: \{chenlanlin, chen\_xiaoming, ning\_ming\}@zju.edu.cn).}
}

\begin{document}

\title{Exploiting On-Orbit Characteristics for Joint Parameter and Channel Tracking in LEO Satellite Communications}

% author names and IEEE memberships
% note positions of commas and nonbreaking spaces ( ~ ) LaTeX will not break
% a structure at a ~ so this keeps an author's name from being broken across
% two lines.
% use \thanks{} to gain access to the first footnote area
% a separate \thanks must be used for each paragraph as LaTeX2e's \thanks
% was not built to handle multiple paragraphs
%

%\author{\IEEEauthorblockN{XXX}

\maketitle

% As a general rule, do not put math, special symbols or citations
% in the abstract or keywords.
\begin{abstract}
In high-dynamic low earth orbit (LEO) satellite communication (SATCOM) systems, frequent channel state information (CSI) acquisition consumes a large number of pilots, which is intolerable in resource-limited SATCOM systems. To tackle this problem, we propose to track the state-dependent parameters including Doppler shift and channel angles, by exploiting the physical and approximate on-orbit mobility characteristics for LEO satellite and ground users (GUs), respectively. As a prerequisite for tracking, we formulate the state evolution models for kinematic (state) parameters of both satellite and GUs, along with the measurement models that describe the relationship between the state-dependent parameters and states. Then the rough estimation of state-dependent parameters is initially conducted, which is used as the measurement results in the subsequent state tracking. Concurrently, the measurement error covariance is predicted based on the formulated Cram$\acute{\text{e}}$r-Rao lower bound (CRLB). Finally, with the extended Kalman filter (EKF)-based state tracking as the bridge, the Doppler shift and channel angles can be further updated and the CSI can also be acquired. Simulation results show that compared to the rough estimation methods, the proposed joint parameter and channel tracking (JPCT) algorithm performs much better in the estimation of state-dependent parameters. Moreover, as to the CSI acquisition, the proposed algorithm can utilize a shorter pilot sequence than benchmark methods under a given estimation accuracy.
\end{abstract}

% Note that keywords are not normally used for peerreview papers.
%\IEEEpeerreviewmaketitle

\section{Introduction}
\IEEEPARstart{S}{atellite} communication (SATCOM) has attracted lots of attention in both academic and industrial fields recently. The satellite used as a base station (BS) or relay has a much higher altitude than ground users (GUs). Thus, it can support wide-coverage communication for remote areas and meet the connection requirements for the growing number of communication devices in various scenarios, such as satellite Internet of Things (IoT) \cite{JChu35,YHe40,ZQu41}. Moreover, SATCOM networks, embedded into the space-air-ground integrated network (SAGIN) \cite{JLiu2,ZLin42}, are expected to realize the prospect of ubiquitous connectivity for 6G wireless networks. Among different satellites classified by orbit height, the low earth orbit (LEO) satellite with altitude from 400 to 2000 km is widely deployed in SATCOM systems (e.g., Starlink and OneWeb projects \cite{LYou3}), thanks to its lower transmission delay and propagation loss than the counterparts \cite{HAl4}. Recently, multiple input multiple output (MIMO) technology has been applied in LEO SATCOM systems, to significantly improve the system performance by exploiting the array gain, which enables direct communication between the satellite and devices. For example, Starlink version 2 will deploy a phased antenna array with the size of 25 $\text{m}^2$ to realize direct communication with cell phones \cite{Bwang33}.

In LEO SATCOM systems with MIMO architectures, channel estimation or parameter estimation is crucial \cite{JHeo5}. On one hand, effective channel estimation ensures the reliability of data detection for communication terminals. On the other hand, the acquisition of accurate channel state information (CSI) helps the beam management at the LEO satellite, which can maximize the performance enhancement brought by MIMO architectures. The research for the channel estimation methods in conventional MIMO terrestrial communications has been well investigated. For example, in millimeter wave (mmWave) MIMO systems, there were compressed sensing (CS)-based methods (e.g., group sparse Bayesian learning (GSBL) \cite{SSri17} and simultaneous orthogonal matching pursuit (SOMP) \cite{JRF18}), super-resolution parameters estimation methods (e.g., the estimating signal parameters via rotational invariance technique (ESPRIT) \cite{ALiao19} and multiple signal classification (MUSIC) \cite{ZGuo20}), tensor decomposition-based estimation methods with low-rank channel characteristics considered \cite{RZhang21, YLin22}, deep learning (DL)-based channel estimation methods \cite{AAbda23, XMa24, JGao36}, and so on. However, the distinctive characteristics of the LEO satellite channel make it difficult to directly apply the existing terrestrial channel estimation methods to the MIMO LEO SATCOM systems. Specifically, the channel estimation in LEO SATCOM systems may meet the following challenges \cite{JHeo5,LYou6}: Firstly, the feedback information of satellite channels is easily outdated due to large transmission delays in SATCOM systems. Secondly, the high mobility of LEO satellites contributes to large Doppler shifts and highly dynamic environments, where high training overhead is required for effective channel estimation or tracking. Thirdly, the large number of MIMO antennas will significantly increase the number of channel coefficients to be estimated, which further increases the channel estimation overhead.

As to the channel aging problem, the authors in \cite{YZhang25} proposed a DL-based channel prediction scheme for LEO SATCOM systems with massive MIMO, where the correlation of changing satellite channels is considered. Specifically, a satellite channel predictor (SCP) was designed by the long short term with memory (LSTM) units, and its effectiveness can be validated by the simulation results. Besides, L. You et al. \cite{LYou3,LYou6} proposed that in the transmission design of MIMO SATCOM systems, the statistical CSI (sCSI) can be used instead of the conventional instantaneous CSI (iCSI). This is due to the fact that the sCSI varies much slower than the iCSI, which can partly alleviate the outdating problem of the estimated channel.

As to the large training overhead problem, several solutions have also been put forward \cite{ALiao7,YLiu8,SHan9,JYu10,BZheng11,TYue12}. On one hand, thanks to the parametric channel expressions, the channel estimation can be transferred to the estimation problem of key channel parameters. Therefore, some researchers have proposed that the prior state information (e.g., position, speed, trajectory, and direction information) of the communication terminals can be leveraged to pre-estimate the state-dependent parameters, including Doppler shift and channel angles \cite{ALiao7,YLiu8,SHan9}. Yet, it is only applicable at the initial stages when the state information is known. On the other hand, since channel tracking requires fewer pilots than conventional channel estimation schemes, various channel tracking schemes have also been considered to tackle this problem. In \cite{JYu10}, a three-dimensional (3D) dynamic turbo approximate message passing (3D-DTAMP) algorithm was proposed, which is based on the developed 3D two-dimensional Markov channel model for unmanned aerial vehicle (UAV)-satellite communications. The authors in \cite{BZheng11} considered the intelligent reflecting surface (IRS)-aided LEO SATCOM systems and proposed a distributed beam tracking method for the angles of both active and passive beams. The work \cite{TYue12} adopted the block-based (BB) channel tracking scheme instead of the conventional symbol-based (SB) channel tracking to save wireless resources, and the optimal block length is studied. Simulation results revealed the superiority of BB channel tracking in high signal-to-noise ratio (SNR) environments. It is worth noting that these tracking schemes all considered direct tracking for key channel parameters including path angles and gains. However, the evolution models of these parameters can not be intuitively obtained in practical applications. For example, the complicated formulation process is needed for probability-based models in \cite{JYu10}, while the linear time-varying models of angles assumed in \cite{BZheng11} and \cite{TYue12} may not suit for the complex wireless environments.

Fortunately, the motion models (including the evolution models of position and velocity) of mobile terminals are effectively exploited in practical applications. The work \cite{ZWei14} and \cite{FLiu16} proposed an extended Kalman filter (EKF)-based tracking scheme for kinematic (state) parameters of communication terminals. Due to the inexplicit relationship between the kinematic parameters and received signal, the authors harnessed the state-dependent parameters as the measurement models, which can be estimated by the matched filter (MF). However, in these works, the measurement error covariance that was essential in the EKF-based tracking was set heuristically, which needs to be further improved.

Motivated by this, we propose to employ and upgrade the framework in works \cite{ZWei14, FLiu16}, where we focus on the tracking of state-dependent parameters (including Doppler shift and channel angles) based on the state tracking, and the measurement error covariance can be effectively predicted in the proposed algorithm. The main contributions of this paper are summarized as follows:

\begin{itemize}
  \item[$\bullet$]By exploiting the physical and approximate on-orbit mobility characteristics of the LEO satellite and GUs, we formulate the state evolution models and measurement models for an LEO SATCOM system with mobile GUs. Under an earth-centered fixed (ECF) coordinate system, the tracking models are developed by utilizing coordinate transformations and geometric relationships between the state-dependent parameters and states.

  \item[$\bullet$]For LEO SATCOM systems, we propose a joint parameter and channel tracking (JPCT) algorithm that is conducted during each block of a transmission frame. Specifically, the proposed algorithm can be divided into five parts: rough Doppler estimation, rough angle estimation, measurement error prediction based on the formulated Cram$\acute{\text{e}}$r-Rao lower bound (CRLB), EKF-based state tracking and parameter update, and CSI acquisition. With the state tracking as the bridge, the accuracy of rough parameter and channel estimation can be significantly improved, while it will not increase pilot overhead and computational complexity orders.

  \item[$\bullet$]Simulation results demonstrate the effectiveness of the proposed JPCT algorithm. As to parameter tracking, the proposed algorithm performs much better than the rough estimation methods, even with the same pilot overhead and computational complexity. As to channel tracking, the proposed algorithm can reduce the pilot overhead for CSI acquisition, thanks to the updated state-dependent parameters. Besides, the effects of key parameters (including the accuracy of the predicted measurement error covariance, prior state evolution noise, etc.) on the proposed algorithm are well investigated.
\end{itemize}

The rest of this paper is organized as follows. Section \ref{Sec2} introduces the LEO SATCOM system model and transmission frame structure. Section \ref{Sec3} formulates the state evolution models and measurement models before the tracking task. Section \ref{Sec4} describes and analyzes the steps of the proposed joint parameter and channel tracking algorithm during each block. In Section \ref{Sec5}, simulation results are given to verify the effectiveness of the proposed algorithm. Finally, the conclusion is presented in Section \ref{Sec6}.

\emph{Notations}: The following notations are adopted throughout this paper: Boldface lowercase letters denote column vectors, boldface uppercase letters denote matrices, and non-bold letters denote scalars, respectively. For a matrix $\boldsymbol{A}$, $\boldsymbol{A}^{T}$, $\boldsymbol{A}^*$, $\boldsymbol{A}^{H}$, $\boldsymbol{A}^{-1}$, $\boldsymbol{A}^\dagger$, $||\boldsymbol{A}||_{F}$, and $\det(\boldsymbol{A})$ denote its transpose, conjugate, conjugate transpose, inverse, pseudo-inverse, Frobenius norm, and determinant. $\boldsymbol{A}[i,j]$, $\boldsymbol{A}[i,:]$, and $\boldsymbol{A}[:,j]$ denote the $(i,j)$-th entry, the $i$-th row, and the $j$-th column of $\boldsymbol{A}$, respectively. For matrices $\boldsymbol{A}_1$ and $\boldsymbol{A}_2$, $\boldsymbol{A}_1\otimes{\boldsymbol{A}_2}$ denotes the Kronecker product of $\boldsymbol{A}_1$ and $\boldsymbol{A}_2$. For a vector $\boldsymbol{\alpha}$, $||\boldsymbol{\alpha}||$ denotes the length of the vector. For vectors $\boldsymbol{\alpha}$ and $\boldsymbol{\beta}$, $\boldsymbol{\alpha} \cdot \boldsymbol{\beta}$ denotes the dot product of $\boldsymbol{\alpha}$ and $\boldsymbol{\beta}$. $\boldsymbol{I}_M$ denotes the $M\times{M}$ identity matrix and $\boldsymbol{0}_{N\times{M}}$ denotes the $N\times{M}$ all-zero matrix. $\mathbb{E}\{\cdot\}$ denotes the statistical expectation, and $\Re\{\cdot\}$ denotes the real part. $\partial (\cdot)$ and $\partial ^2(\cdot)$ denote operations for the first- and second-order partial derivatives. $\text{diag}\{ \mathcal{A}\}$ denotes a diagonal matrix with main diagonal elements from set $\mathcal{A}$. $\mathcal{CN}(\boldsymbol{\mu},\boldsymbol{Q})$ denotes the complex Gaussian distribution with mean vector $\boldsymbol{\mu}$ and covariance matrix $\boldsymbol{Q}$.
\begin{figure*}
  \centering
  \includegraphics[width=0.7\textwidth]{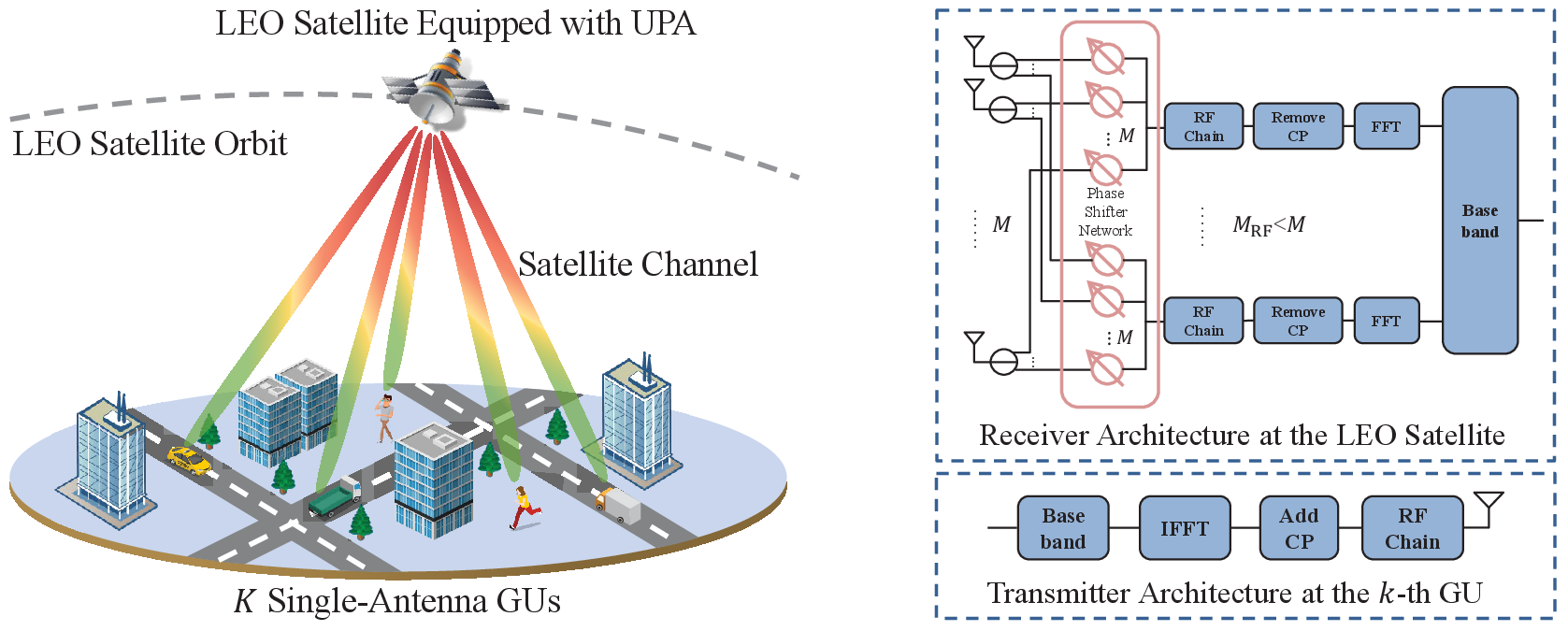}
  \caption{Illustration and transceiver architecture of a multi-user LEO SATCOM system.}\label{SystemModel}
\end{figure*}

\section{System Model}\label{Sec2}
As shown in Fig. \ref{SystemModel}, consider a multi-user LEO SATCOM system, where an LEO satellite equipped with $M$ antennas serves $K$ single-antenna mobile GUs. To balance performance gain and power consumption, a hybrid beamforming architecture is employed at the satellite \cite{LYou3}. Specifically, the hybrid architecture conducts digital beamforming at the baseband and analog beamforming based on a phase shifter network with $M_\text{RF}$ $(M_\text{RF}<M)$ radio frequency (RF) chains at the RF. Moreover, the orthogonal frequency division multiplexing (OFDM) technology with $N_\text{sc}$ subcarriers is adopted to resist the multi-path fading. From Fig. \ref{SystemModel}, it can be observed that the OFDM technology is applied by performing the inverse fast Fourier transform (IFFT) and adding the cyclic prefix (CP) to the baseband signal at the transmitter. Meanwhile, to recover the baseband signal, the inverse operations including removing the CP and performing the fast Fourier transform (FFT) are conducted at the receiver. Besides, to avoid inter-user interference, each subcarrier channel is assigned to one GU at most.

In order to realize efficient multi-user LEO SATCOM on fast time-varying channels caused by the movement of both satellite and GUs, it is desired to perform parameter and channel tracking. In the following, we introduce the frame structure, channel model, and signal model for joint parameter and channel tracking.

\subsection{Frame Structure}
\begin{figure}
  \centering
  \includegraphics[width=0.45\textwidth]{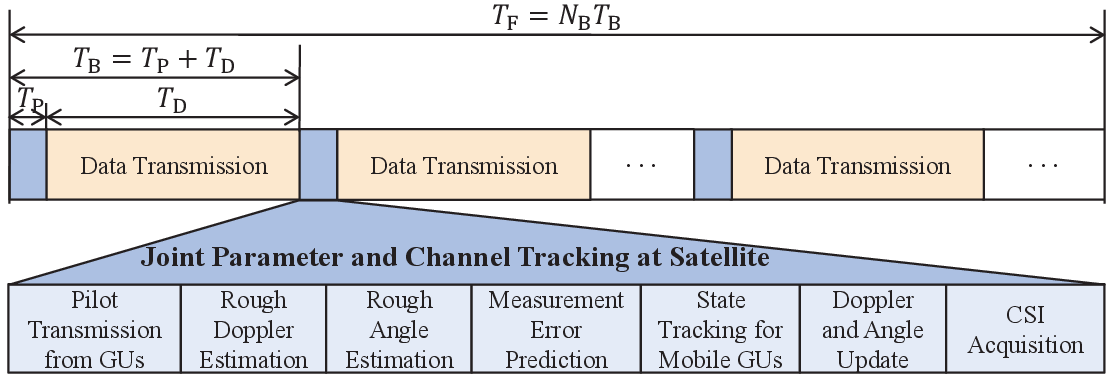}
  \caption{Frame structure of the proposed joint parameter and channel tracking algorithm.}\label{FrameStructure}
\end{figure}

To decrease the computational amount and signaling overhead due to frequent channel tracking, we perform block-based tracking instead of symbol-based tracking in this paper \cite{TYue12}. A specific block-based frame structure is given in Fig. \ref{FrameStructure}, where a transmission frame consists of $N_\text{B}$ blocks, and each block includes a pilot component and a data component. In Fig. \ref{FrameStructure}, it can be observed that the duration of one block is $T_\text{B}=T_\text{P}+T_\text{D} = N_\text{ofdm}T_\text{sym}$, where $N_\text{ofdm}$ denotes the number of OFDM symbols in one block, $T_\text{sym}$ denotes the duration of one OFDM symbol, $T_\text{P} = N_\text{P}T_\text{sym}$ is the duration of pilot component in one block, $T_\text{D} = (N_\text{ofdm}-N_\text{P}) T_\text{sym}$ is the duration of data component in one block, and $N_\text{P}$ denotes the number of OFDM symbols that are used as pilots. With $N_\text{B}$ blocks, the duration of one frame can be calculated as $T_\text{F} = N_\text{B}T_\text{B}$. The duration of one block is limited, and thus we assume that the system parameters remain unchanged during each block. At the beginning of each block, with the pilot sent from the GUs, the LEO satellite conducts joint parameter and channel tracking. Specifically, the rough Doppler shift and channel angles are firstly estimated. Subsequently, the measurement error covariance of the rough estimation is predicted based on the formulated CRLB. Then, with the constructed tracking models, the state tracking for mobile GUs and satellite can be achieved, and thus the state-dependent parameters can be updated. Finally, the complete CSI can be acquired with the updated parameters. After that, the parameters\footnote{The updated Doppler shift and channel angles can aid for sensing in the integrated sensing and communication (ISAC) systems. Besides, as a substitute for the complete channel, the channel angles can also be used for designing beamforming schemes, with the advantage of lower overhead \cite{XHu39}.} and CSI are used for data transmission in the rest of the block.

% \vspace{-0.4cm}
\subsection{Channel Model}
Consider the direct communication scenario in LEO SATCOM systems, the working frequency band is set as L/S/C band in this paper. According to the commonly used geometric parametric channel model in LEO SATCOM systems with the OFDM modulation, the $M$-dimensional discrete channel at the $n$-th block and the $m$-th subcarrier for the $k$-th GU can be written as \cite{LYou6,ALiao7}
\begin{equation}\label{equa27}
\boldsymbol{h}_k^m(n)= \sum_{l=0}^{L_k-1}\tilde{g}_{k,l}(n) \boldsymbol{a}_{k,l}(n) e^{-j2\pi (f_c+m\triangle f)\tau_{k,l}(n)},
\end{equation}
where $L_k$ denotes the number of propagation paths for the $k$-th GU, $\tilde{g}_{k,l}(n)$, $\boldsymbol{a}_{k,l}(n)$ and $\tau_{k,l}(n)$ are the equivalent complex channel gain, array response vector at the LEO satellite, and propagation delay over the $l$-th path, respectively. $f_c$ denotes the carrier frequency and $\triangle f$ is the interval of subcarriers.

Furthermore, with the block index $n$ neglected for simplicity, $\tilde{g}_{k,l}$ can be divided into two components: the line of sight (LOS) component with $l=0$ and the non-line of sight (NLOS) component with $l = 1, \cdots, L_k-1$, which can be expressed as \cite{MYing13}
\begin{equation}\label{equa25}
\begin{split}
\tilde{g}_{k,0} &= \gamma_k\sqrt{\frac{\lambda_k\beta_k}{\lambda_k+1}}g_{k,0}, \\
\tilde{g}_{k,l} &= \gamma_k\sqrt{\frac{\beta_k}{\lambda_k+1}}\sqrt{\frac{1}{L_k-1}} g_{k,l},\\
l &= 1, \cdots, L_k-1,
\end{split}
\end{equation}
where $g_{k,l}, l = 0, \cdots, L_k-1$ are the actual channel gains for the $k$-th GU over the $l$-th path, $\gamma_k$ is the receive antenna gain, $\lambda_k$ is the Rician factor, and $\beta_k$ is the large-scale fading factor, which can be described as
\begin{equation}\label{equa2}
  \beta_k = \left(\frac{c}{4\pi f_cD_k}\right)^2 \cdot \frac{G_k}{\kappa B_wT},
\end{equation}
where $c$ denotes the light speed, $D_k$ denotes the LOS propagation distance between the satellite and the $k$-th GU, and $(\frac{c}{4\pi f_cD_k})^2$ is the free space loss. In addition, $G_k$ is the transmit antenna gain, $\kappa$ is Boltzmann's constant, $B_w$ is the channel bandwidth, and $T$ is the temperature of the received noise.

Meanwhile, it is reasonably assumed that all the propagation paths of a certain GU have the same angle of arrival (AoA), since the altitude of the LEO satellite is much higher than that of the scatters distributed around the GU \cite{LYou6}, namely $\boldsymbol{a}_{k,0} = \boldsymbol{a}_{k,l} = \boldsymbol{a}_{k}$. Moreover, we consider a uniform planar array (UPA) with half wavelength antenna distances and $M=M_xM_y$ antennas mounted on the satellite, where $M_x$ and $M_y$ denote the number of antennas on the $x$-axis and $y$-axis, respectively. Thereby, the array response vector can be expressed as \cite{ZWei14}
\begin{equation}\label{equa5}
\begin{split}
\boldsymbol{a}_k = \boldsymbol{a}_{k,x}\otimes \boldsymbol{a}_{k,y} &\in \mathbb{C}^{M\times 1}, \\
\boldsymbol{a}_{k,x} = 1/\sqrt{M_x}\cdot [1, e^{j\pi\bar{\theta}_{k,x}}, &\dots, e^{j\pi(M_x-1)\bar{\theta}_{k,x}}]^T, \\
\boldsymbol{a}_{k,y} = 1/\sqrt{M_y}\cdot [1, e^{j\pi\bar{\theta}_{k,y}}, &\dots, e^{j\pi (M_y-1)\bar{\theta}_{k,y}}]^T,
\end{split}
\end{equation}
where $\bar{\theta}_ {k,x}=\sin(\theta_{k,\text{A}})\sin(\theta_{k,\text{E}})$ and $\bar{\theta}_{k,y} = \cos(\theta_{k,\text{E}})$ represent the virtual AoAs with $\theta_{k,\text{E}}\in [0, \pi/2)$ and $\theta_{k,\text{A}}\in [-\pi/2,\pi/2)$ being the elevation and azimuth of AoAs for the $k$-th GU, respectively. With $\boldsymbol{a}_{k,l}(n) = \boldsymbol{a}(\theta_{k,\text{E}}(n), \theta_{k,\text{A}}(n))$, the discrete channel can be rewritten as
\begin{equation}\label{equa26}
\begin{split}
&\boldsymbol{h}_k^m(n)= \\ &\sum_{l=0}^{L_k-1}\tilde{g}_{k,l}(n) \boldsymbol{a}(\theta_{k,\text{E}}(n), \theta_{k,\text{A}}(n)) e^{-j2\pi (f_c+m\triangle f)\tau_{k,l}(n)},
\end{split}
\end{equation}
where the channel parameters (including $\tilde{g}_{k,l}(n)$, $\theta_{k,\text{E}}(n)$, $\theta_{k,\text{A}}(n)$, and $\tau_{k,l}(n)$) are time-varying over blocks.

\subsection{Signal Model}
Without loss of generality, assume that the $m$-th subcarrier channel is assigned to the $k$-th GU. During the $n$-th block, the $k$-th GU transmits a pilot sequence with the length of $N_\text{P}$ over the $m$-th subcarrier in the continuous $N_\text{P}$ OFDM symbols. Thereby, at the RF end of the satellite, the received signal at the $b$-th OFDM symbol and $m$-th subcarrier for the $k$-th GU $\boldsymbol{r}_k^m(n,b) \in \mathbb{C}^{M_\text{RF}\times 1}$ can be expressed as
\begin{equation}\label{equa29}
\begin{split}
&\boldsymbol{r}_k^m(n,b) = \\
&\sqrt{P_k}\boldsymbol{W}_k \boldsymbol{h}_k^m(n) s_k^m(n,b) e^{j2\pi u_k(n) b T_\text{sym}} + \boldsymbol{W}_k{\boldsymbol{n}}_k^m(n,b),
\end{split}
\end{equation}
where $n\in\{0, 1, \cdots, N_\text{B}-1\}$, $b\in \{0, 1, \cdots, N_\text{P}-1\}$, $m\in \{0, 1, \cdots, N_\text{sc}-1\}$, $P_k$ denotes the transmit power for the $k$-th GU,  $\boldsymbol{W}_k\in \mathbb{C}^{M_\text{RF} \times M}$ denotes the analog combiner matrix at the RF end of the LEO satellite, which has constant modulus constraint with $||\boldsymbol{W}_k||_{F}^2 = 1$, $s_k^m(n,b)$ denotes the pilot signal that satisfies $|s_k^m(n,b)|^2 = 1$ (such as Zadoff-Chu (ZC) sequence \cite{MHuang29}), $u_k(n)$ is the Doppler shift associated with the $k$-th GU that is time-varying over blocks, and $T_\text{sym}$ is the duration of one OFDM symbol. ${\boldsymbol{n}}_k^m(n,b) \sim \mathcal{CN}(\boldsymbol{0}_{M\times 1},\sigma_n^2\boldsymbol{I}_{M})$ is the additive white Gaussian noise (AWGN) with zero mean vector and covariance matrix $\sigma_n^2\boldsymbol{I}_{M}$.

Furthermore, multiplying the received signal $\boldsymbol{r}_k^m(n,b)$ by $[s_k^m(n,b)]^*$ and
substituting the expression of channel $ \boldsymbol{h}_k^m(n)$ into (\ref{equa29}), the received signal can be transformed as (\ref{equa4}) at the top of next page, where ${\bar{\boldsymbol{n}}}_k^m(n,b) = \boldsymbol{W}_k{\boldsymbol{n}}_k^m(n,b) [s_k^m(n,b)]^* $. Note that Doppler shift $u_k(n)$, elevation angle $\theta_{k,\text{E}}(n)$, and azimuth angle $\theta_{k,\text{A}}(n)$ in (\ref{equa4}) are the state-dependent parameters that need to be tracked in this paper.

\begin{figure*}
\begin{equation}\label{equa4}
\bar{\boldsymbol{r}}_k^m(n,b) = \boldsymbol{r}_k^m(n,b)[s_k^m(n,b)]^* = \sqrt{P_k}\boldsymbol{W}_k \boldsymbol{a}(\theta_{k,\text{E}}(n), \theta_{k,\text{A}}(n)) e^{j2\pi u_k(n) b T_\text{sym}} \sum_{l=0}^{L_k-1}\tilde{g}_{k,l}(n) e^{-j2\pi (f_c+m\triangle f)\tau_{k,l}(n)} + {\bar{\boldsymbol{n}}}_k^m(n,b).
\end{equation}
\hrule
\end{figure*}

\section{Formulation of Tracking Models}\label{Sec3}
In this section, we will present the state evolution models and measurement models according to the characteristics of LEO SATCOM systems, which is an essential prerequisite for parameter tracking. As previously assumed, the states (position and velocity of the satellite and GUs) and state-dependent parameters (Doppler shift and channel angles) are invariant during each block but vary between blocks. Then, we will formulate the tracking models during the $n$-th block.

\subsection{Coordinate System and Assumptions}
\begin{figure*}
  \centering
  \includegraphics[width=0.8\textwidth]{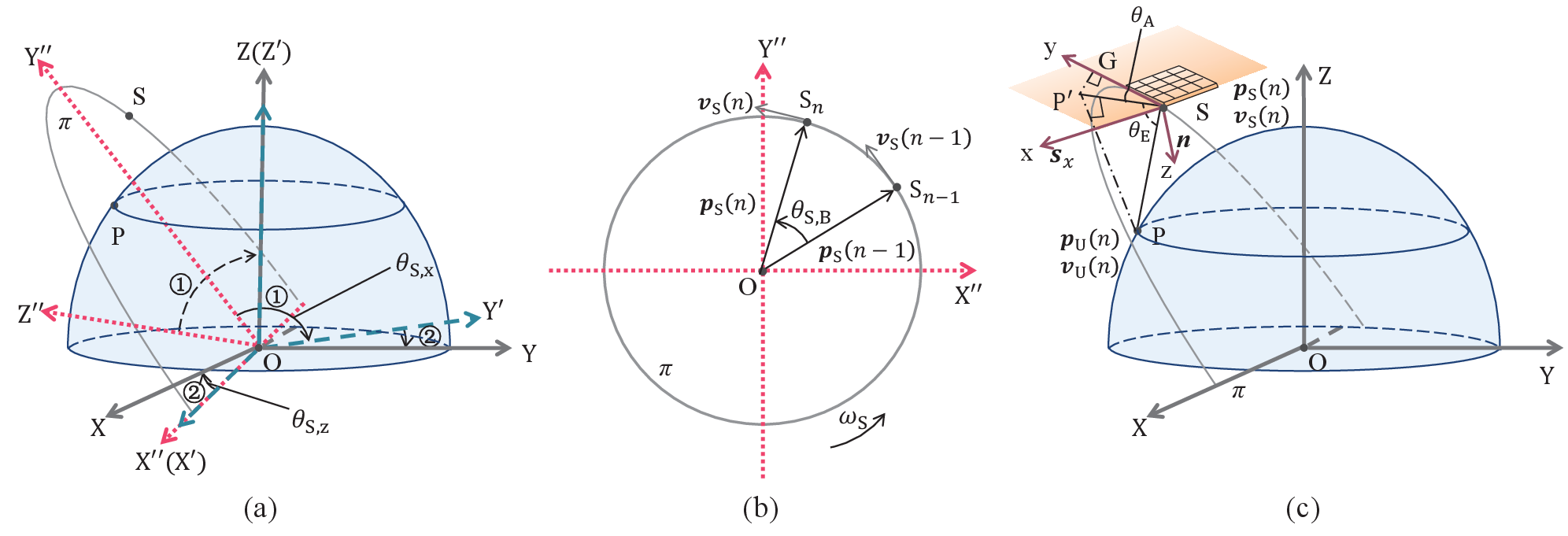}
  \caption{Geometric relationship diagram for LEO satellite communication based on the ECF coordinate system: (a) coordinate system rotation; (b) rotation on the satellite orbit plane; and (c) presentation for measurement model formulation.}\label{GR}
\end{figure*}

For notational simplicity, we focus on the construction of models between the satellite and the $k$-th GU, where the index $k$ is omitted. As shown in Fig. \ref{GR}, under the unified ECF coordinate system XYZ, the origin point O is the center of the earth, point S denotes the position of the LEO satellite, and point P denotes the position of the $k$-th GU. Moreover, in the $n$-th block, the position and velocity vectors for satellite and GU in three-dimensional space can be denoted by $\boldsymbol{p}_\text{S}(n)$, $\boldsymbol{p}_\text{U}(n)$, $\boldsymbol{v}_\text{S}(n)$, and $\boldsymbol{v}_\text{U}(n)$, respectively.

Due to the high-speed mobility of LEO satellites, the communication duration for one moving cycle of the satellite is relatively short, which is generally called by ``visibility window''. Therefore, during a communication period, the satellite orbit can be approximated as a large arc \cite{IAli15}, and it is reasonably assumed that the LEO satellite moves uniformly in a circular orbit with the same center of the earth. Meanwhile, we assume the approximate on-orbit characteristics for mobile GUs, especially for those scenarios or applications where the users move with a specific regularity.\footnote{For example, the vehicle or pedestrian traveling usually follows a specific route based on a certain criterion such as shortest commuting time; in the field of military communication, guided missiles typically have specific ballistic trajectories for a determined target \cite{KMoon37}; in high-speed railway applications, the train trajectory planning can be determined under a given timetable and running routes \cite{XHY38}. For these scenarios or applications, the user's trajectory can be approximately expanded to a moving orbit for the mobile user, which greatly helps the formulation of tracking models.} Furthermore, within the period of a ``visibility window'', it can be approximately assumed that the GU moves uniformly on the surface of the earth along a circular orbit with the same center of the earth, and random noise may exist due to modeling errors \cite{ZWei14}.

\subsection{State Evolution Model}
In general, the state-dependent parameters, i.e., Doppler shift and AoAs, vary with the mobility of the LEO satellite and GUs. In other words, the state-dependent parameters depend on the system states, including the position and velocity vectors for satellite and GUs. Specifically, the Doppler shift depends on the velocity and position of satellite and GUs, while the AoAs depend on the position of satellite and GUs (see Section III. C for detailed formulations of the expressions between state-dependent parameters and states). To track these state-dependent parameters, we should build the evolution models of system states, which refer to the relationship between state vectors of the $n$-th block and the $(n-1)$-th block. According to the previous assumption, let the LEO satellite move in a general circular orbit with the same center of the earth, and the constant angular velocity of the satellite is denoted by $\omega_\text{S}$. Since the LEO satellite moves on the two-dimensional orbit plane, it is challenging to formulate the evolution models of the position and velocity vectors in the three-dimensional space directly. Responding to this, we propose a modeling method based on the coordinate system rotation principles. For convenience, we define two kinds of rotation matrices as follows:
\begin{equation}\label{equa22}
\begin{cases}
  \boldsymbol{P}_{\text{\uppercase\expandafter{\romannumeral1}}}(\theta) = \begin{bmatrix}
  \cos\theta & -\sin\theta & 0 \\
  \sin\theta & \cos\theta & 0 \\
  0 & 0 & 1
\end{bmatrix} \\
  \boldsymbol{P}_{\text{\uppercase\expandafter{\romannumeral2}}}(\theta) = \begin{bmatrix}
  1 & 0 & 0 \\
  0 & \cos\theta & -\sin\theta \\
  0 & \sin\theta & \cos\theta
\end{bmatrix}
\end{cases},
\end{equation}
where $\theta$ is the angle of rotation. According to the characteristics of rotation matrices, the inverse matrix of a rotation matrix is equal to its transpose matrix, which can be described by
\begin{equation}\label{equa79} \boldsymbol{P}_{\text{\uppercase\expandafter{\romannumeral1}}}^{-1}(\theta) = \boldsymbol{P}_{\text{\uppercase\expandafter{\romannumeral1}}}^T(\theta), \boldsymbol{P}_{\text{\uppercase\expandafter{\romannumeral2}}}^{-1}(\theta) = \boldsymbol{P}_{\text{\uppercase\expandafter{\romannumeral2}}}^{T}(\theta) .
\end{equation}

As shown in Fig. \ref{GR}(a), let the intersection between satellite orbit plane $\pi$ and XOY plane as the $\text{X}''$-axis, the orbit plane $\pi$ as the $\text{X}''\text{OY}''$ plane, then the $\text{X}''\text{Y}''\text{Z}''$ coordinate system can be established. Afterward, the rotation process from coordinate system $\text{X}''\text{Y}''\text{Z}''$ to XYZ can be summarized as follows: 1) Rotate the system $\text{X}''\text{Y}''\text{Z}''$ clockwise around the $\text{X}''$-axis by $\theta_{\text{S},x}$, where $\theta_{\text{S},x}$ is the angle between the satellite orbit plane $\pi$ and the XOY plane. Thereby, the intermediate coordinate system $\text{X}'\text{Y}'\text{Z}'$ can be obtained, where the $\text{X}'$-axis coincides with the $\text{X}''$-axis and the $\text{Z}'$-axis coincides with the Z-axis. 2) To obtain the XYZ coordinate system, rotate the system $\text{X}'\text{Y}'\text{Z}'$ clockwise around the $\text{Z}'$-axis by $\theta_{\text{S}, z}$, where $\theta_{\text{S}, z}$ denotes the angle between
the X-axis and the $\text{X}''$-axis. If any point/vector is denoted by $[x,y,z]^T$ under the system XYZ or $[x'',y'',z'']^T$ under the system $\text{X}''\text{Y}''\text{Z}''$, then the transformation relationship between these two coordinate systems of the same point/vector can be given by
\begin{equation}\label{equa80}
[x,y,z]^T = \boldsymbol{P}_{\text{S},z}\boldsymbol{P}_{\text{S},x} [x'',y'',z'']^T,
\end{equation}
where $\boldsymbol{P}_{\text{S},z} = \boldsymbol{P}_{\text{\uppercase\expandafter{\romannumeral1}}}(\theta_{\text{S},z})$ and $\boldsymbol{P}_{\text{S},x} = \boldsymbol{P}_{\text{\uppercase\expandafter{\romannumeral2}}}(\theta_{\text{S},x})$.
Furthermore, we can also obtain the following relationship
\begin{equation}\label{equa37}
\begin{split}
[x'',y'',z'']^T &= (\boldsymbol{P}_{\text{S},z}\boldsymbol{P}_{\text{S},x})^{-1} [x,y,z]^T \\
&=\boldsymbol{P}_{\text{S},x}^{-1} \boldsymbol{P}_{\text{S},z}^{-1}[x,y,z]^T \\
&=\boldsymbol{P}_{\text{S},x}^{T} \boldsymbol{P}_{\text{S},z}^{T}[x,y,z]^T,
\end{split}
\end{equation}
where the third equation is due to the characteristics of rotation matrices shown in (\ref{equa79}).

Subsequently, we consider the state evolution of the satellite over blocks on the satellite orbit plane $\pi$. As shown in Fig. \ref{GR}(b), assume that the LEO satellite moves counterclockwise\footnote{If the satellite moves clockwise, the formulation of evolution models is similar by letting $\theta_\text{S,B} = -\theta_\text{S,B}$.} with the angular velocity $\omega_\text{S}$, then the rotation angle between blocks is $\theta_\text{S,B} = \omega_\text{S}T_\text{B}$, where $T_\text{B}$ is the duration of one block. Thus, through simple geometric derivations, the evolution expressions of position and velocity vectors under the $\text{X}''\text{Y}''\text{Z}''$ coordinate system can be obtained as
\begin{equation}\label{equa81}
\boldsymbol{p}_\text{S}''(n) = \begin{bmatrix}
x''(n)\\y''(n)\\z''(n)\end{bmatrix} = \boldsymbol{P}_\text{S,B}\begin{bmatrix}x''(n-1)\\ y''(n-1)\\ z''(n-1)\end{bmatrix} = \boldsymbol{P}_\text{S,B}\boldsymbol{p}_\text{S}''(n-1),
\end{equation}
\begin{equation}\label{equa82}
\boldsymbol{v}_\text{S}''(n) = \begin{bmatrix}
v_x''(n)\\v_y''(n)\\v_z''(n)\end{bmatrix} = \boldsymbol{P}_\text{S,B}\begin{bmatrix}v_x''(n-1)\\ v_y''(n-1)\\ v_z''(n-1)\end{bmatrix} = \boldsymbol{P}_\text{S,B}\boldsymbol{v}_\text{S}''(n-1),
\end{equation}
where $\boldsymbol{P}_\text{S,B} = \boldsymbol{P}_{\text{\uppercase\expandafter{\romannumeral1}}} (\theta_\text{S,B})$.

As to the evolution model for position vector, substitute (\ref{equa37}) into (\ref{equa81}), and we can obtain
\begin{equation}\label{equa38}
\begin{split}
\boldsymbol{p}_\text{S}''(n) &= (\boldsymbol{P}_{\text{S},z}\boldsymbol{P}_{\text{S},x})^{-1} \boldsymbol{p}_\text{S}(n)= \boldsymbol{P}_\text{S,B}\boldsymbol{p}_\text{S}''(n-1) \\ &=\boldsymbol{P}_\text{S,B} \boldsymbol{P}_{\text{S},x}^{T} \boldsymbol{P}_{\text{S},z}^{T}\boldsymbol{p}_\text{S}(n-1).
\end{split}
\end{equation}
Therefore, the evolution model for position vector of the satellite can be expressed as
\begin{equation}\label{equa39}
\boldsymbol{p}_\text{S}(n) = \boldsymbol{P}_{\text{S},z}\boldsymbol{P}_{\text{S},x} \boldsymbol{P}_\text{S,B} \boldsymbol{P}_{\text{S},x}^{T} \boldsymbol{P}_{\text{S},z}^{T}\boldsymbol{p}_\text{S}(n-1) = \boldsymbol{F}_\text{S} \boldsymbol{p}_\text{S}(n-1).
\end{equation}
Similarly, the evolution model for velocity vector of the satellite can be given as
\begin{equation}\label{equa41}
\boldsymbol{v}_\text{S}(n) = \boldsymbol{F}_\text{S} \boldsymbol{v}_\text{S}(n-1),
\end{equation}
where $\boldsymbol{F}_\text{S} = \boldsymbol{P}_{\text{S},z}\boldsymbol{P}_{\text{S}, x} \boldsymbol{P}_{\text{S,B}}\boldsymbol{P}_{\text{S}, x}^T \boldsymbol{P}_{\text{S}, z}^T$.

According to the expressions in (\ref{equa39}) and (\ref{equa41}), the state evolution models for the LEO satellite can be constructed as
\begin{equation}\label{equa8}
\begin{split}
\boldsymbol{q}_\text{S}(n) &=
\begin{bmatrix}
  \boldsymbol{p}_\text{S}(n) \\ \boldsymbol{v}_\text{S}(n)
\end{bmatrix} =
\begin{bmatrix}
  \boldsymbol{F}_\text{S} & \boldsymbol{0} \\
  \boldsymbol{0} & \boldsymbol{F}_\text{S}
\end{bmatrix}\cdot
\begin{bmatrix}
  \boldsymbol{p}_\text{S}(n-1) \\
  \boldsymbol{v}_\text{S}(n-1)
\end{bmatrix} \\
&= \tilde{\boldsymbol{F}}_\text{S}\boldsymbol{q}_\text{S}(n-1),
\end{split}
\end{equation}
where $\boldsymbol{q}_\text{S}(n)$ denotes the state vector of the satellite in the $n$-th block.

Considering the previous discussions, let the GU move on the surface of the earth along a general circular orbit with the same center of the earth, and the constant angular velocity of the GU is denoted by $\omega_\text{U}$. Then the formulation of state evolution models for GUs is similar to that for the LEO satellite. Meanwhile, with the random evolution noise considered, the state evolution models for the GU can be constructed as
\begin{equation}\label{equa9}
  \begin{split}
     \boldsymbol{q}_{\text{U}}(n) &=
     \begin{bmatrix}
       \boldsymbol{p}_{\text{U}}(n) \\
       \boldsymbol{v}_{\text{U}}(n)
     \end{bmatrix} = \begin{bmatrix}
     \boldsymbol{F}_{\text{U}} & \boldsymbol{0} \\
  \boldsymbol{0} & \boldsymbol{F}_{\text{U}}
  \end{bmatrix}\cdot \begin{bmatrix}
       \boldsymbol{p}_{\text{U}}(n-1) \\
       \boldsymbol{v}_{\text{U}}(n-1)
     \end{bmatrix}+\begin{bmatrix}
       \boldsymbol{w}_{p}(n) \\
       \boldsymbol{w}_{v}(n)
     \end{bmatrix} \\
       &= \tilde{\boldsymbol{F}}_{\text{U}}\boldsymbol{q}_{\text{U}} (n-1)+\boldsymbol{w}_{\text{U}}(n),
  \end{split}
\end{equation}
where $\boldsymbol{q}_{\text{U}}(n)$ denotes the state vector of the GU in the $n$-th block, $\boldsymbol{F}_{\text{U}} = \boldsymbol{P}_{\text{U},z}\boldsymbol{P}_{\text{U}, x} \boldsymbol{P}_{\text{U,B}}\boldsymbol{P}_{\text{U}, x}^T \boldsymbol{P}_{\text{U}, z}^T$, $\boldsymbol{P}_{\text{U},z} = \boldsymbol{P}_{\text{\uppercase\expandafter{\romannumeral1}}}(\theta_{\text{U},z})$, $\boldsymbol{P}_{\text{U},x} = \boldsymbol{P}_{\text{\uppercase\expandafter{\romannumeral2}}}(\theta_{\text{U},x})$, $\boldsymbol{P}_\text{U,B} = \boldsymbol{P}_{\text{\uppercase\expandafter{\romannumeral1}}} (\theta_\text{U,B})$, $\theta_{\text{U},z}$ denotes the angle between the X-axis and the intersection line that is between the orbit plane of the GU and the XOY plane, $\theta_{\text{U},x}$ is the angle between the orbit plane of the GU and the XOY plane, $\theta_{\text{U,B}} = \omega_\text{U}T_\text{B}$ is the angle that the GU rotates on the orbit plane within one block (which is positive with counterclockwise rotation and negative with clockwise rotation). ${\boldsymbol{w}}_{\text{U}}(n) \sim \mathcal{CN}(\boldsymbol{0}, \boldsymbol{Q}_\text{U})$ is the state evolution noise of the GU, which obeys complex Gaussian distribution with zero mean and covariance matrix $\boldsymbol{Q}_\text{U} = \text{diag}\{\sigma_{x_\text{u}}^2, \sigma_{y_\text{u}}^2, \sigma_{z_\text{u}}^2, \sigma_{v_x}^2, \sigma_{v_y}^2, \sigma_{v_z}^2\}$, where $\sigma_{x_\text{u}}^2$, $\sigma_{y_\text{u}}^2$, $\sigma_{z_\text{u}}^2$, $\sigma_{v_x}^2$, $\sigma_{v_y}^2$ and $\sigma_{v_z}^2$ are evolution noise variances of $\boldsymbol{q}_{\text{U}}(n)$.

\subsection{Measurement Model}
Measurement models refer to the mapping expressions from state vectors of the LEO satellite and the GU to the state-dependent parameters including Doppler shift $u$, elevation angle $\theta_\text{E}$, and azimuth angle $\theta_\text{A}$. As shown in Fig. \ref{GR}(c), a UPA array is mounted on the LEO satellite, and then through geometric relationship derivations, the measurement models can be formulated as
\begin{equation}\label{equa10}
\begin{split}
  u(n) &= \frac{-[\boldsymbol{v}_\text{S}(n)-\boldsymbol{v}_{\text{U}}(n)] \cdot [\boldsymbol{p}_\text{S}(n)-\boldsymbol{p}_{\text{U}}(n)])}{\lambda_c ||\boldsymbol{p}_\text{S}(n)-\boldsymbol{p}_{\text{U}}(n)||} \\
  &= g_u(\boldsymbol{q}_\text{S}(n), \boldsymbol{q}_{
  \text{U}}(n)),
\end{split}
\end{equation}
\begin{equation}\label{equa11}
\begin{split}
  \theta_{\text{E}}(n) &= \arcsin\{\frac{|\boldsymbol{n} \cdot [\boldsymbol{p}_\text{S}(n)-\boldsymbol{p}_{\text{U}}(n)]|}{||\boldsymbol{n}|| \cdot ||\boldsymbol{p}_\text{S}(n)-\boldsymbol{p}_{\text{U}}(n)||}\} \\
  &= g_\text{E}(\boldsymbol{q}_\text{S}(n), \boldsymbol{q}_{
  \text{U}}(n)),
\end{split}
\end{equation}
\begin{equation}\label{equa12}
\begin{split}
  \theta_{\text{A}}(n) &= \frac{\pi}{2} - \arccos\{\frac{\vec{\text{SP}'}\cdot \boldsymbol{s}_x}{\|\vec{\text{SP}'}\|\cdot \|\boldsymbol{s}_x\|}\} \\
  &= g_\text{A}(\boldsymbol{q}_\text{S}(n), \boldsymbol{q}_{
  \text{U}}(n)),
\end{split}
\end{equation}
where $\lambda_c$ is the carrier wavelength, $\boldsymbol{n}$ is the normal vector of the UPA plane which is an expansion of the UPA array on the satellite, $\vec{\text{SP}'}$ is the vector from the satellite position point S to the projection point $\text{P}'$ of the GU position point P on the UPA plane, $\boldsymbol{s}_x$ is the direction vector of the x-axis in another coordinate system xyz where the point S is set as the origin point and the UPA plane is set as the xSy plane. $g_u(\cdot)$, $g_\text{E}(\cdot)$, and $g_\text{A}(\cdot)$ are the mapping functions from the state vectors of the satellite and the GU to these state-dependent parameters, which can be noted as $\boldsymbol{g}(\cdot) = [g_u(\cdot), g_\text{E}(\cdot), g_\text{A}(\cdot)]^T$. Note that the direction vector $\boldsymbol{s}_x$ and normal vector $\boldsymbol{n}$ can be decided by the position of the satellite and UPA configuration. Meanwhile, the projection point $\text{P}'$ can be easily obtained with the normal vector of the UPA plane, the position of GU, and the position of the satellite. For example, with the normal vector $\boldsymbol{n} = [n_x, n_y, n_z]^T$, the position of satellite $\boldsymbol{p}_\text{S} = [p_\text{S}^x, p_\text{S}^y, p_\text{S}^z]^T$, and the position of GU $\boldsymbol{p}_\text{U} = [p_\text{U}^x, p_\text{U}^y, p_\text{U}^z]^T$, the elements of $\text{P}'$ $\{{p'}_\text{U}^x, {p'}_\text{U}^y, {p'}_\text{U}^z\}$ can be obtained by simple spatial geometric operations as
\begin{equation}\label{equa72}
\begin{cases}
{p'}_\text{U}^x &= \frac{1}{||\boldsymbol{n}||^2} \left[(n_y^2+n_z^2)p_\text{U}^x - n_x(n_yp_\text{U}^y+n_zp_\text{U}^z + C_\text{S}) \right] \\
{p'}_\text{U}^y &= \frac{1}{||\boldsymbol{n}||^2} \left[(n_x^2+n_z^2)p_\text{U}^y - n_y(n_xp_\text{U}^x+n_zp_\text{U}^z + C_\text{S}) \right] \\
{p'}_\text{U}^z &= \frac{1}{||\boldsymbol{n}||^2} \left[(n_x^2+n_y^2)p_\text{U}^z - n_z(n_xp_\text{U}^x+n_yp_\text{U}^y + C_\text{S}) \right]
\end{cases},
\end{equation}
where $C_\text{S} = -n_xp_\text{S}^x-n_yp_\text{S}^y-n_zp_\text{S}^z$.

For the $k$-th GU, with $\boldsymbol{z}_k(n) = [u_k(n), \theta_{k,\text{E}}(n), \theta_{k,\text{A}}(n)]^T$ as the target parameter vector, the overall measurement model can be obtained as
\begin{equation}\label{equa30}
\boldsymbol{z}_k(n) = \boldsymbol{g}(\boldsymbol{q}_\text{S}(n), \boldsymbol{q}_{k,
  \text{U}}(n)).
\end{equation}

\section{Design of Joint Parameter and Channel Tracking Algorithm}\label{Sec4}
In this section, with the constructed state evolution models and measurement models, we design a joint parameter and channel tracking algorithm for LEO SATCOM systems. According to Fig. \ref{FrameStructure}, during the tracking process in each block, the LEO satellite receives the pilots from $K$ GUs firstly, then the rough Doppler and angle estimation\footnote{Different from the works \cite{ZWei14, FLiu16}, we focus on the estimation accuracy of the state-dependent parameters. Therefore, we adopt the commonly used Doppler and angles estimation methods for MIMO parametric channel models, instead of the MF method, which may not guarantee the estimation performance.} are carried out for each GU. With the rough estimates, variances of the measurement error can be predicted based on the derived CRLB. Subsequently, according to the state evolution models and measurement models developed in Section \ref{Sec3}, the state tracking for LEO satellite and mobile GUs can be realized through an EKF process \cite{ZWei14,FLiu16}. Finally, with the tracked state vectors, the target parameters can be updated through the mapping expressions of the measurement models, and the complete channel can be acquired in the meantime.

As to the $k$-th GU, the pilot sequence of length $N_\text{P}$ is transmitted over the ${m}$-th subcarrier for parameter estimation and tracking.
For simplicity, we neglect the index of subcarrier channel $m$ that corresponds to the $k$-th GU in the following presentations. Accordingly, the received signal can be noted as $\{\bar{\boldsymbol{r}}_k(n,b)\}_{b=0}^{N_\text{P}-1}$ with the same expression of (\ref{equa4}). In the following, we design the joint parameter and channel tracking algorithm based on the received signal.

\subsection{Rough Doppler Estimation}
From the expression in (\ref{equa4}), it can be observed that the Doppler shift is at the exponential phase and can be roughly estimated by the ESPRIT algorithm \cite{ALiao7, RRoy26}. Specifically, based on the received signal $\{\bar{\boldsymbol{r}}_k(n,b)\}_{b=0}^{N_\text{P}-1}$, we can define the following two matrices with the same dimension of $(N_\text{P}-1)\times M_\text{RF}$
\begin{equation}\label{equa6}
\begin{split}
\boldsymbol{Y}_{k,1}(n) &= [\bar{\boldsymbol{r}}_k(n,0), \bar{\boldsymbol{r}}_k(n,1), \cdots, \bar{\boldsymbol{r}}_k(n,N_\text{P}-2)]^T, \\
\boldsymbol{Y}_{k,2}(n) &= [\bar{\boldsymbol{r}}_k(n,1), \bar{\boldsymbol{r}}_k(n,2), \cdots, \bar{\boldsymbol{r}}_k(n,N_\text{P}-1)]^T.
\end{split}
\end{equation}
Afterward, the auto-correlation and cross-correlation matrices can be respectively computed as
\begin{equation}\label{equa78}
\begin{split}
\boldsymbol{R}_{k,11}(n) &= \frac{1}{(N_\text{P}-1)M_\text{RF}}\boldsymbol{Y}_{k,1}(n)\boldsymbol{Y}_{k,1}^H(n), \\ \boldsymbol{R}_{k,12}(n) &= \frac{1}{(N_\text{P}-1)M_\text{RF}}\boldsymbol{Y}_{k,1}(n)\boldsymbol{Y}_{k,2}^H(n).
\end{split}
\end{equation}
Through eigenvalue decomposition on $\boldsymbol{R}_{k,11}(n)$, its minimum eigenvalue $\lambda_{k,\text{min}}(n)$ can be acquired. Subsequently, the revised correlation matrices can be rewritten as
\begin{equation}\label{equa1}
\begin{split}
\boldsymbol{\bar{R}}_{k,11}(n) &= \boldsymbol{R}_{k,11}(n) - \lambda_{k,\text{min}}(n)\boldsymbol{I}_{N_\text{P}-1}, \\ \boldsymbol{\bar{R}}_{k,12}(n) &= \boldsymbol{R}_{k,12}(n) - \lambda_{k,\text{min}}(n)\boldsymbol{V}, \\
\boldsymbol{V} &= \begin{bmatrix} \boldsymbol{0}_{1\times (N_\text{P}-2)} & 0 \\ \boldsymbol{I}_{N_\text{P}-2} & \boldsymbol{0}_{(N_\text{P}-2)\times 1} \end{bmatrix}.
\end{split}
\end{equation}
Calculate the eigenvalue $\lambda_k(n)$ after the generalized eigenvalue decomposition of $\{\boldsymbol{\bar{R}}_{k,11}(n), \boldsymbol{\bar{R}}_{k,12}(n)\}$, and then the roughly estimated Doppler shift of the $k$-th GU in the $n$-th block can be obtained as
\begin{equation}\label{equa13}
\hat{u}_k^{\text{RE}}(n) = \frac{\ln(\lambda_k(n))}{j2\pi T_\text{sym}}.
\end{equation}

\subsection{Rough Angle Estimation}
With the discrete presentation for AoAs, the received signal can be approximated as
\begin{equation}\label{equa28}
\begin{split}
\bar{\boldsymbol{r}}_k(n,b) \approx \sqrt{P_k}\boldsymbol{W}_k \boldsymbol{A} \boldsymbol{x}_k(n,b) + {\bar{\boldsymbol{n}}}_k(n,b), \\
\boldsymbol{A} = [\boldsymbol{a}(\phi_{\text{E},0}, \phi_{\text{A},0}), \cdots, \boldsymbol{a}(\phi_{\text{E},0}, \phi_{\text{A},N_\text{A}-1}), \\
\cdots, \boldsymbol{a}(\phi_{\text{E},N_\text{E}-1}, \phi_{\text{A},N_\text{A}-1})] \in \mathbb{C}^{M\times G}, \\
\boldsymbol{x}_k(n,b) = [0, \cdots, C_{k,\boldsymbol{z}}(n,b) ,\cdots, 0]^T \in \mathbb{C}^{G\times 1},
\end{split}
\end{equation}
where $\boldsymbol{A}$ denotes dictionary matrix consists of discrete elevation $\phi_{\text{E},i} \in \{0, \frac{\pi}{2N_\text{E}}, \cdots, \frac{\pi}{2N_\text{E}}(N_\text{E}-1)\}$ and azimuth angles $\phi_{\text{A},j} \in \{-\frac{\pi}{2}, -\frac{\pi}{2}+ \frac{\pi}{N_\text{A}}, \cdots, -\frac{\pi}{2}+ \frac{\pi}{N_\text{A}}(N_\text{A}-1)\}$, $N_\text{E}$ and $N_\text{A}$ denote the resolution for elevation angle and azimuth angle respectively, and $G = N_\text{E}N_\text{A}(G>M)$ is the number of grid points. $\boldsymbol{x}_k(n,b)$ is the sparse vector with one non-zero element $C_{k,\boldsymbol{z}}(n,b) = e^{j2\pi u_k(n) b T_\text{sym}} \sum_{l=0}^{L_k-1}\tilde{g}_{k,l}(n) e^{-j2\pi (f_c+{m}\triangle f)\tau_{k,l}(n)}$. The position of the non-zero element coincides with the nearest discrete AoAs to the real AoAs in the dictionary matrix, which is also called sparse support. Therefore, the angle estimation problem can be transferred into the acquisition of the sparse support and can be tackled by the SOMP algorithm \cite{JRF18}. The SOMP algorithm is a classic CS method, its principle is to maximize the relevance between the received signal and the sensing matrix. Thus, the sparse support can be obtained as
\begin{equation}\label{equa24}
\{i^*, j^*\} = \mathop{\arg\max}\limits_{i,j}\sum_{b=0}^{N_\text{P}-1} \left| \bigl\{\boldsymbol{W}_k\boldsymbol{A}[:,g]\bigr\}^H \bar{\boldsymbol{r}}_k(n,b) \right|,
\end{equation}
where $g = iN_\text{A}+j$, $i\in \{0, 1, \cdots, N_\text{E}-1\}$, and $j \in \{0, 1, \cdots, N_\text{A}-1\}$. With the sparse support acquired, the rough estimates for AoAs can be given as
\begin{equation}\label{equa83}
\hat{\theta}_{k,\text{E}}^{\text{RE}}(n) = \phi_{\text{E},i^*}, \hat{\theta}_{k,\text{A}}^{\text{RE}}(n) = \phi_{\text{A},j^*}.
\end{equation}

Besides, to obtain better estimation performance, we propose a design scheme for the combiner matrix as follows:
In the first block, the combiner matrix $\boldsymbol{W}_k$ is set as the discrete Fourier transform (DFT) matrix with the dimension of $M_\text{RF}\times M$. In the following blocks, with the updated AoAs of the previous block $\{\hat{{\theta}}_{k, \text{E}}(n-1), \hat{{\theta}}_{k, \text{A}}(n-1)\}$, the first row of the combiner matrix is designed as
\begin{equation}\label{equa32}
\boldsymbol{W}_k[0,:] = \boldsymbol{a}^H(\hat{{\theta}}_{k, \text{E}}(n-1), \hat{{\theta}}_{k, \text{A}}(n-1)),
\end{equation}
which can maximize the received signal power \cite{CLin30}.

Based on the above discussions, we can obtain the rough estimates for the target parameter vector $\hat{\boldsymbol{z}}_k^{\text{RE}}(n) = [\hat{u}_k^{\text{RE}}(n), \hat{\theta}_{k,\text{E}}^{\text{RE}}(n), \hat{\theta}_{k,\text{A}}^{\text{RE}}(n)]^T$.

\subsection{Measurement Error Prediction Based on CRLB}
In the proposed algorithm, the rough estimates are used as the measurement results in the tracking process, and thus the measurement error prediction is to predict the error variances of the rough estimation methods. Since the real results of the parameters to be estimated are usually unavailable, it is difficult to predict the error variances of the rough estimation methods. Fortunately, the CRLB is a classic tool for evaluating the performance of parameter estimators, which characterizes the lower bound of the error variances of the estimators \cite{JRF18,ALiao7, FLiu16}. Therefore, we propose to utilize the CRLB to predict the error variances of the aforementioned rough estimation methods\footnote{If some real results of the parameters to be estimated are available, we can simply calculate the error variances of the real and estimated data, or deploy the machine learning (ML) algorithms to predict the measurement error variances.}. In the following, the formulation of the CRLB for rough Doppler and angle estimation will be provided respectively, and after that, the measurement error prediction is conducted. Note that the CRLB will be developed in the $n$-th block and for the $k$-th GU, and thus, we further neglect the index $k$ and $n$ for simplicity.

\vspace{0.2cm}
\subsubsection{CRLB for Rough Doppler Estimation}
With the received signal as (\ref{equa29}), the log-likelihood function of the received signal ${\boldsymbol{r}}$ and Doppler shift $u$ can be written as (\ref{equa33}) at the top of next page, where $\boldsymbol{\Sigma} = \boldsymbol{W}\boldsymbol{W}^H \in \mathbb{C}^{M_\text{RF}\times M_\text{RF}}$.

Prior to calculating the Fisher information matrix (FIM), the expectation for second derivatives of the log-likelihood function needs to be obtained. Accordingly, the second derivatives of $\ln p({\boldsymbol{r}};u)$ on the Doppler shift $u$ can be firstly calculated as (\ref{equa61}) at the top of next page, where we neglect the details of derivative calculations for saving space.

\begin{figure*}
\begin{equation}\label{equa33}
\resizebox{0.85\hsize}{!}{$
\begin{split}
\ln p({\boldsymbol{r}};u) = -M_\text{RF}N_\text{P} \ln(\pi \det(\sigma_n^2\boldsymbol{\Sigma})) &- \frac{1}{\sigma_n^2} \sum_{b=0}^{N_\text{P}-1} \biggl\{\bigl[ {\boldsymbol{r}}(b) - \sqrt{P}\boldsymbol{W} \boldsymbol{a}(\theta_{\text{E}}, \theta_{\text{A}})s(b) e^{j2\pi u b T_\text{sym}} \sum_{l=0}^{L-1}\tilde{g}_{l} e^{-j2\pi (f_c+{m}\triangle f)\tau_{l}}\bigr]^H \\
&\cdot\boldsymbol{\Sigma}^{-1}\cdot \bigl[ {\boldsymbol{r}}(b) - \sqrt{P}\boldsymbol{W} \boldsymbol{a}(\theta_{\text{E}}, \theta_{\text{A}}) s(b) e^{j2\pi u b T_\text{sym}} \sum_{l=0}^{L-1}\tilde{g}_{l} e^{-j2\pi (f_c+{m}\triangle f)\tau_{l}}\bigr]\biggr\}.
\end{split}
$}
\end{equation}

\begin{equation}\label{equa61}
\resizebox{0.85\hsize}{!}{$
\frac{\partial^2 \ln p({\boldsymbol{r}};u)}{\partial u^2} =
-\frac{8\pi^2 \sqrt{P}T_\text{sym}^2}{\sigma_n^2} \sum_{b=0}^{N_\text{P}-1}\Re\bigl\{ {\boldsymbol{r}}^H(b) \boldsymbol{\Sigma}^{-1} \boldsymbol{W} \boldsymbol{a}(\theta_{\text{E}}, \theta_{\text{A}})\cdot b^2 s(b) e^{j2\pi u b T_\text{sym}} \sum_{l=0}^{L-1}\tilde{g}_{l} e^{-j2\pi (f_c+{m}\triangle f)\tau_{l}} \bigr\}.
$}
\end{equation}
\end{figure*}

From the expression in (\ref{equa61}), it is obvious that the development of $\mathbb{E}\left\{\frac{\partial^2 \ln p({\boldsymbol{r}};u)}{\partial u^2}\right\}$ refers to the calculation of $\mathbb{E}\{{\boldsymbol{r}}(b)\}$, which can be easily given as
\begin{equation}\label{equa34}
\resizebox{0.97\hsize}{!}{$
\mathbb{E}\{{\boldsymbol{r}}(b)\} =
\sqrt{P}\boldsymbol{W} \boldsymbol{a}(\theta_{\text{E}}, \theta_{\text{A}}) s(b) e^{j2\pi u b T_\text{sym}} \sum_{l=0}^{L-1}\tilde{g}_{l} e^{-j2\pi (f_c+{m}\triangle f)\tau_{l}},
$}
\end{equation}
due to the white Gaussian noise in the expression of the received signal as (\ref{equa4}). Afterward, the FIM $I_u$ can be calculated as (\ref{equa68}) at the top of next page, where
\begin{figure*}
\begin{equation}\label{equa68}
\resizebox{0.7\hsize}{!}{$
  I_u = -\mathbb{E}\left\{\frac{\partial^2 \ln p({\boldsymbol{r}};u)}{\partial u^2}\right\} = \frac{4\pi^2 }{3\sigma_n^2} P T_\text{sym}^2 N_\text{P}(N_\text{P}-1)(2N_\text{P}-1)\bigl\| \boldsymbol{\Sigma}^{-\frac{1}{2}}\boldsymbol{W} \boldsymbol{a}(\theta_{\text{E}}, \theta_{\text{A}})\bigr\|^2 |\tilde{C}|^2.
  $}
\end{equation}
\hrule
\end{figure*}
\begin{equation}\label{equa74}
  \tilde{C} = \sum_{l=0}^{L-1}\tilde{g}_{l} e^{-j2\pi (f_c+{m}\triangle f)\tau_{l}}.
\end{equation}

Based on the above discussions, the mean square error (MSE) of the Doppler estimation can be bounded by
\begin{equation}\label{equa31}
\mathbb{E}\left\{(\hat{u}-u)^2
\right\}\geq I_u^{-1} \triangleq \text{CRLB}_u(\boldsymbol{\theta}, \tilde{C}),
\end{equation}
where the CRLB of Doppler estimation depends on the AoAs $\boldsymbol{\theta} = [\theta_\text{E}, \theta_\text{A}]^T$ and the summation term $\tilde{C}$.

\vspace{0.2cm}
\subsubsection{CRLB for Rough Angle Estimation}
\begin{figure*}
\begin{equation}\label{equa69}
\resizebox{0.55\hsize}{!}{$
  \boldsymbol{I}_\theta[0,0] = -\mathbb{E}\left\{\frac{\partial^2 \ln p({\boldsymbol{r}};\boldsymbol{\theta})}{\partial \theta_\text{E}^2}\right\} = \frac{2PN_\text{P}}{\sigma_n^2}\bigl\|\dot{\boldsymbol{a}}_\text{E} \boldsymbol{W}^T (\boldsymbol{\Sigma}^{-\frac{1}{2}})^T\bigr\|^2 |\tilde{C}|^2,
  $}
\end{equation}
\begin{equation}\label{equa70}
\resizebox{0.55\hsize}{!}{$
  \boldsymbol{I}_\theta[1,1] = -\mathbb{E}\left\{\frac{\partial^2 \ln p({\boldsymbol{r}};\boldsymbol{\theta})}{\partial \theta_\text{A}^2}\right\} = \frac{2PN_\text{P}}{\sigma_n^2} \bigl\|\dot{\boldsymbol{a}}_\text{A} \boldsymbol{W}^T (\boldsymbol{\Sigma}^{-\frac{1}{2}})^T\bigr\|^2 |\tilde{C}|^2,
  $}
\end{equation}
\begin{equation}\label{equa73}
\resizebox{0.7\hsize}{!}{$
  \boldsymbol{I}_\theta[0,1] = \boldsymbol{I}_\theta[1,0] = -\mathbb{E}\left\{\frac{\partial^2 \ln p({\boldsymbol{r}};\boldsymbol{\theta})}{\partial \theta_\text{A} \partial \theta_\text{E}}\right\} = \frac{2PN_\text{P}}{\sigma_n^2} \Re\bigl\{\dot{\boldsymbol{a}}_\text{E} \boldsymbol{W}^T (\boldsymbol{\Sigma}^{-1})^* \boldsymbol{W}^* \dot{\boldsymbol{a}}_\text{A}^H \bigr\} |\tilde{C}|^2.
  $}
\end{equation}
\hrule
\end{figure*}

As for the rough angle estimation, the received signal forms are the same with rough Doppler estimation as (\ref{equa4}), where $b\in\{0,1,\cdots, N_\text{P}-1\}$. Therefore, similar to the formulation of the CRLB for Doppler estimation, the elements of FIM $\boldsymbol{I}_\theta$ for angle estimation can be developed as (\ref{equa69})-(\ref{equa73}) at the top of next page, where
\begin{equation}\label{equa35}
\begin{split}
\dot{\boldsymbol{a}}_{\text{E}}= \frac{\partial(\boldsymbol{a}(\theta_\text{E}, \theta_\text{A})) }{\partial \theta_\text{E}}&\in\mathbb{C}^{1\times M}, \\ \dot{\boldsymbol{a}}_{\text{A}}= \frac{\partial(\boldsymbol{a}(\theta_\text{E}, \theta_\text{A})) }{\partial \theta_\text{A}}&\in\mathbb{C}^{1\times M}.
\end{split}
\end{equation}

Accordingly, the MSE of the angle estimation can be bounded by
\begin{equation}\label{equa44}
\mathbb{E}\left\{(\boldsymbol{\hat{\theta}} - \boldsymbol{\theta})(\boldsymbol{\hat{\theta}} - \boldsymbol{\theta})^H
\right\}\succeq \boldsymbol{I}_\theta^{-1} \triangleq \boldsymbol{J},
\end{equation}
\begin{equation}\label{equa65}
\mathbb{E}\left\{(\hat{\theta}_\text{E}-\theta_\text{E})^2
\right\}\geq \boldsymbol{J}[0,0] \triangleq \text{CRLB}_\text{E}(\boldsymbol{\theta}, \tilde{C}),
\end{equation}
\begin{equation}\label{equa66}
\mathbb{E}\left\{(\hat{\theta}_\text{A}-\theta_\text{A})^2
\right\}\geq \boldsymbol{J}[1,1] \triangleq \text{CRLB}_\text{A}(\boldsymbol{\theta}, \tilde{C}),
\end{equation}
where the CRLB of angle estimation depends on the AoAs $\boldsymbol{\theta} = [\theta_\text{E}, \theta_\text{A}]^T$ and the summation term $\tilde{C}$.

\vspace{0.2cm}
\subsubsection{Measurement Error Prediction}
Without loss of generality, the estimation errors can be approximated by the Gaussian noise with zero mean and covariance matrix $\boldsymbol{Q}_{z} = \text{diag}\{\sigma_{u}^2, \sigma_{\text{E}}^2, \sigma_{\text{A}}^2\}$, where $\sigma_{u}^2$, $\sigma_{\text{E}}^2$, and $\sigma_{\text{A}}^2$ are the variances of the error caused by the rough estimation of Doppler shift, elevation angle, and azimuth angle. In the $n$-th block and for the $k$-th GU, we attempt to predict the measurement error based on the formulated CRLB. From the expressions in (\ref{equa31}), (\ref{equa65}), and (\ref{equa66}), it can be observed that the formulated CRLB depends on the AoAs $\boldsymbol{\theta}$ and the summation term $\tilde{C}$. To guarantee the accuracy of predicted CRLB, the CRLB-related variables can be obtained by utilizing rough estimates ($n=0$) or tracking results of the previous block ($n>0$). Thus, the available estimation results are $\boldsymbol{\hat{z}}^{\text{CRLB}} = [\hat{u}^{\text{CRLB}}, \hat{\theta}_\text{E}^{\text{CRLB}}, \hat{\theta}_\text{A}^{\text{CRLB}}]^T$, which can be given as
\begin{equation}\label{equa75}
\boldsymbol{\hat{z}}^{\text{CRLB}} =
\begin{cases}
\boldsymbol{\hat{z}}^{\text{RE}}, &n=0 \\
\boldsymbol{\hat{z}}(n-1), &n>0
\end{cases},
\end{equation}
where $\boldsymbol{\hat{z}}^{\text{RE}} = [\hat{u}^{\text{RE}}, \hat{\theta}_\text{E}^{\text{RE}}, \hat{\theta}_\text{A}^{\text{RE}}]^T$ and $\boldsymbol{\hat{z}}(n-1)=[\hat{u}(n-1), \hat{\theta}_\text{E}(n-1), \hat{\theta}_\text{A}(n-1)]^T$.

With the estimation results $\boldsymbol{\hat{z}}^{\text{CRLB}}$, the AoAs can be obtained by $\boldsymbol{\hat{\theta}}^{\text{CRLB}} = [\hat{\theta}_\text{E}^{\text{CRLB}}, \hat{\theta}_\text{A}^{\text{CRLB}}]^T $. Moreover, the summation term can be predicted by the least square (LS) method \cite{BZheng11}. Specifically, since the summation term is uncorrelated to the symbol index $b$, it can be predicted by applying the LS method over $\{\boldsymbol{r}(b)\}_{b=0}^{N_\text{P}-1}$ and making expectations as follows
\begin{equation}\label{equa36}
\resizebox{0.98\hsize}{!}{$
\hat{\tilde{C}}=\frac{1}{N_\text{P}}\sum_{b=0}^{N_\text{P}-1} \left[\sqrt{P}\boldsymbol{W} \boldsymbol{a}(\hat{\theta}_\text{E}^{\text{CRLB}}, \hat{\theta}_\text{A}^{\text{CRLB}})\right]^\dagger {\boldsymbol{r}}(b) e^{-j2\pi \hat{u}^{\text{CRLB}} b T_\text{sym}}.
$}
\end{equation}

Accordingly, the variances of the measurement error can be predicted as
\begin{equation}\label{equa71}
\begin{cases}
  \hat{\sigma}_{u}^2 &= \text{CRLB}_u(\boldsymbol{\hat{\theta}}^{\text{CRLB}}, \hat{\tilde{C}}) \\
  \hat{\sigma}_{\text{E}}^2 &= \text{CRLB}_\text{E}(\boldsymbol{\hat{\theta}}^{\text{CRLB}}, \hat{\tilde{C}}) \\
  \hat{\sigma}_{\text{A}}^2 &= \text{CRLB}_\text{A}(\boldsymbol{\hat{\theta}}^{\text{CRLB}}, \hat{\tilde{C}})
\end{cases}.
\end{equation}

Therefore, in the $n$-th block, the predicted covariance matrix of the measurement error for the $k$-th GU can be obtained by $\hat{\boldsymbol{Q}}_{k,z}(n) = \text{diag}\{\hat{\sigma}_{k,u}^2(n), \hat{\sigma}_{k,\text{E}}^2(n), \hat{\sigma}_{k,\text{A}}^2(n)\}$, where $\hat{\sigma}_{k,u}^2(n)$, $\hat{\sigma}_{k,\text{E}}^2(n)$, and $\hat{\sigma}_{k,\text{A}}^2(n)$ are predicted in the $n$-th block and for the $k$-th GU, based on the expressions in (\ref{equa71}).

\subsection{State Tracking and Parameter Update}
\begin{algorithm}[t]
\caption{Joint Parameter and Channel Tracking Algorithm}
\hspace*{0.02in} {\bf Input:}
State parameters of previous block $\hat{\boldsymbol{q}}_\text{S}(n-1)$, $\hat{\boldsymbol{q}}_{k,\text{U}}(n-1)$, and $\boldsymbol{C}(n-1)$, state transition matrix $\tilde{\boldsymbol{F}}_\text{S}$ and $\tilde{\boldsymbol{F}}_{k,\text{U}}$, innovation noise covariance $\boldsymbol{Q}_\text{U}$, channel noise variance $\sigma_n^2$, length of pilot signal $N_\text{P}$, index of occupied subcarrier channel ${m}$ for the $k$-th GU, updated parameters of previous block $\{\hat{u}_k(n-1), \hat{{\theta}}_{k, \text{E}}(n-1), \hat{{\theta}}_{k, \text{A}}(n-1)\}$
\begin{algorithmic}[1]
\State \% {\bf Rough Estimation}
\State Design combiner matrix $\boldsymbol{W}_k$ based on updated AoAs of previous block $\{\hat{{\theta}}_{k, \text{E}}(n-1), \hat{{\theta}}_{k, \text{A}}(n-1)\}$ using (\ref{equa32})
\State Obtain received signal $\bar{\boldsymbol{r}}_k(n,b)$ with combiner matrix $\boldsymbol{W}_k$ as (\ref{equa4}), for $b\in\{0, 1, \cdots, N_\text{P}-1\}$
\State Estimate Doppler shift $\hat{u}_k^{\text{RE}}(n)$ roughly by ESPRIT algorithm using (\ref{equa6})-(\ref{equa13})
\State Estimate angles $\hat{\theta}_{k,\text{E}}^{\text{RE}}(n)$ and $\hat{\theta}_{k,\text{A}}^{\text{RE}}(n)$ roughly by SOMP algorithm using (\ref{equa24})-(\ref{equa83})
\State \% {\bf Measurement Error Prediction}
\State Obtain the CRLB-related variables $\boldsymbol{\hat{z}}_k^{\text{CRLB}}(n)$, $\boldsymbol{\hat{\theta}}_k^{\text{CRLB}}(n)$, and $\hat{\tilde{C}}$ using (\ref{equa75}) and (\ref{equa36})
\State Predict $\text{CRLB}_u(\boldsymbol{\hat{\theta}}_k^{\text{CRLB}}(n), \hat{\tilde{C}})$ for Doppler shift using (\ref{equa68}) and (\ref{equa31})
\State Predict $\text{CRLB}_\text{E}(\boldsymbol{\hat{\theta}}_k^{\text{CRLB}}(n), \hat{\tilde{C}})$ and  $\text{CRLB}_\text{A}(\boldsymbol{\hat{\theta}}_k^{\text{CRLB}}(n), \hat{\tilde{C}})$ for AoAs using (\ref{equa69})-(\ref{equa73}) and (\ref{equa44})-(\ref{equa66})
\State Estimate measurement error covariance $\hat{\boldsymbol{Q}}_{k,z}(n)$ based on the predicted CRLBs using (\ref{equa71})
\State \% {\bf State Tracking}
\State Predict system state parameters $\hat{\boldsymbol{q}}_\text{S}(n)$, $\hat{\boldsymbol{q}}_{k,\text{U}}(n|n-1)$ and $\boldsymbol{C}(n|n-1)$ using (\ref{equa14}), (\ref{equa15}), and (\ref{equa16})
\State Obtain the initial parameter estimates $\hat{\boldsymbol{z}}_k(n|n-1) = \boldsymbol{g}(\hat{\boldsymbol{q}}_\text{S}(n), \hat{\boldsymbol{q}}_{k,\text{U}}(n|n-1))$
\State Calculate the Kalman gain based on predicted measurement error covariance $\hat{\boldsymbol{Q}}_{k,z}(n)$ using (\ref{equa19}) and (\ref{equa20})
\State Obtain updated states $\hat{\boldsymbol{q}}_{k,\text{U}}(n)$ and $\boldsymbol{C}(n)$ based on the Kalman gain and results from the rough estimation $\hat{\boldsymbol{z}}_k^{\text{RE}}(n) = [\hat{u}_k^{\text{RE}}(n), \hat{\theta}_{k,\text{E}}^{\text{RE}}(n), \hat{\theta}_{k,\text{A}}^{\text{RE}}(n)]^T$ using (\ref{equa17}) and (\ref{equa18})
\State \% {\bf Parameter Update}
\State Obtain updated system parameters $\hat{{u}}_k(n)$, $\hat{{\theta}}_{k, \text{E}}(n)$, and $\hat{{\theta}}_{k, \text{A}}(n)$ using (\ref{equa21})
\State \% {\bf CSI Acquisition}
\State Obtain CSI $\boldsymbol{\hat{h}}_k(n)$ by the LS algorithm using (\ref{equa40})
\end{algorithmic}
\hspace*{0.02in} {\bf Output:}
$\hat{\boldsymbol{z}}_k(n) = [\hat{{u}}_k(n), \hat{{\theta}}_{k, \text{E}}(n), \hat{{\theta}}_{k, \text{A}}(n)]^T$, $\hat{\boldsymbol{q}}_\text{S}(n)$, $\hat{\boldsymbol{q}}_{k,\text{U}}(n)$, $\boldsymbol{C}(n)$, and $\boldsymbol{\hat{h}}_k(n)$
\end{algorithm}

In the practical scenarios, the distribution or evolution models for the target parameters (including Doppler shift and channel angles) are difficult to be acquired. Fortunately, by exploiting the channel characteristics of the LEO SATCOM systems, there exists mapping relationships between the target parameters and states of communication terminals, which enables the target parameters to be state-dependent. Hence, by using the state tracking as the bridge, the estimated state-dependent parameters can be further updated. It is worth noting that compared to the rough estimation methods presented previously, the proposed joint parameter and channel tracking algorithm can achieve better estimation performance under the same pilot overhead. Specifically, with the state evolution models developed in (\ref{equa8})-(\ref{equa9}) and measurement models in (\ref{equa10})-(\ref{equa12}), the parameter tracking can be conducted by the EKF process\footnote{To realize the EKF-based tracking, the initial position and velocity of satellite and GUs are needed at the satellite end. On one hand, the initial position and velocity of the satellite can be extracted from its ephemeris information. On the other hand, the initial position and velocity of GUs can be acquired locally first. For example, the initial location of GUs on the earth can be obtained by the global navigation satellite system (GNSS), and the initial velocity of GUs can be calculated by exploiting speed measurement devices. Subsequently, each GU can transmit its initial information to the LEO satellite, as a prerequisite for the following tracking.}.

In each block, based on the estimated information in the previous block $\hat{\boldsymbol{q}}_\text{S}(n-1)$, $\hat{\boldsymbol{q}}_{k,\text{U}}(n-1)$, and $\boldsymbol{C}(n-1)$, the states can be firstly predicted as
\begin{equation}\label{equa14}
  \hat{\boldsymbol{q}}_\text{S}(n) = \tilde{\boldsymbol{F}}_\text{S}\hat{\boldsymbol{q}}_\text{S}(n-1),
\end{equation}
\begin{equation}\label{equa15}
 \hat{\boldsymbol{q}}_{k,\text{U}}(n|n-1) = \tilde{\boldsymbol{F}}_{k,\text{U}}\hat{\boldsymbol{q}}_{k,\text{U}}(n-1),
\end{equation}
\begin{equation}\label{equa16}
  \boldsymbol{C}(n|n-1) = \tilde{\boldsymbol{F}}_{k,\text{U}}\boldsymbol{C}(n-1)\tilde{\boldsymbol{F}}_{k,\text{U}}^H + \boldsymbol{Q}_\text{U},
\end{equation}
where $\boldsymbol{C}(n-1)$ is the covariance matrix of EKF in the $(n-1)$-th block. Therefore, the initial parameter estimates are $\hat{\boldsymbol{z}}_k(n|n-1) = \boldsymbol{g}(\hat{\boldsymbol{q}}_\text{S}(n), \hat{\boldsymbol{q}}_{k,\text{U}}(n|n-1))$, as defined in (\ref{equa30}).

Afterward, the states can be updated according to the estimated results $\hat{\boldsymbol{z}}_k^{\text{RE}}(n)$ from the rough estimation previously, where the covariance matrix of measurement error can be predicted by $\hat{\boldsymbol{Q}}_{k,z}(n)$ based on the formulated CRLB. Mathematically, the updated states of the GUs and covariance matrix can be derived respectively as
\begin{equation}\label{equa17}
\hat{\boldsymbol{q}}_{k,\text{U}}(n) = \hat{\boldsymbol{q}}_{k,\text{U}}(n|n-1) + \boldsymbol{K}(n)[\hat{\boldsymbol{z}}_k^{\text{RE}}(n) - \hat{\boldsymbol{z}}_k(n|n-1)],
\end{equation}
\begin{equation}\label{equa18}
\boldsymbol{C}(n) = [\boldsymbol{I} - \boldsymbol{K}(n)\boldsymbol{G}(n)]\boldsymbol{C}(n|n-1),
\end{equation}
where $\boldsymbol{K}(n)$ is the Kalman gain and $\boldsymbol{G}(n)$ is the Jacobin matrix, which are respectively given by
\begin{equation}\label{equa19}
\resizebox{0.97\hsize}{!}{$
\boldsymbol{K}(n) = \boldsymbol{C}(n|n-1)\boldsymbol{G}^H(n) [\boldsymbol{G}(n)\boldsymbol{C}(n|n-1)\boldsymbol{G}^H(n)+\hat{ \boldsymbol{Q}}_{k,z}(n)]^{-1},
$}
\end{equation}
\begin{equation}\label{equa20}
\resizebox{0.98\hsize}{!}{$
\boldsymbol{G}(n) = \left.\begin{bmatrix}
\frac{\partial{u_k(n)}}{\partial{x_{k,\text{u}}(n)}} & \frac{\partial{u_k(n)}}{\partial{y_{k,\text{u}}(n)}} & \frac{\partial{u_k(n)}}{\partial{z_{k,\text{u}}(n)}} & \frac{\partial{u_k(n)}}{\partial{v_{k,\text{u}, x}(n)}} & \frac{\partial{u_k(n)}}{\partial{v_{k,\text{u}, y}(n)}} & \frac{\partial{u_k(n)}}{\partial{v_{k,\text{u}, z}(n)}} \\
\frac{\partial{\theta_{k,\text{E}}(n)}}{\partial{x_{k,\text{u}}(n)}} & \frac{\partial{\theta_{k,\text{E}}(n)}}{\partial{y_{k,\text{u}}(n)}} & \frac{\partial{\theta_{k,\text{E}}(n)}}{\partial{z_{k,\text{u}}(n)}} & \frac{\partial{\theta_{k,\text{E}}(n)}}{\partial{v_{k,\text{u}, x}(n)}} & \frac{\partial{\theta_{k,\text{E}}(n)}}{\partial{v_{k,\text{u}, y}(n)}} & \frac{\partial{\theta_{k,\text{E}}(n)}}{\partial{v_{k,\text{u}, z}(n)}} \\
\frac{\partial \theta_{k, \text{A}}(n)}{\partial{x_{k,\text{u}}(n)}} & \frac{\partial \theta_{k, \text{A}}(n)}{\partial{y_{k,\text{u}}(n)}} & \frac{\partial \theta_{k, \text{A}}(n)}{\partial{z_{k,\text{u}}(n)}} & \frac{\partial \theta_{k, \text{A}}(n)}{\partial{v_{k,\text{u}, x}(n)}} & \frac{\partial \theta_{k, \text{A}}(n)}{\partial{v_{k,\text{u}, y}(n)}} & \frac{\partial \theta_{k, \text{A}}(n)}{\partial{v_{k,\text{u}, z}(n)}}
\end{bmatrix}\right|_{\boldsymbol{q}_{k, \text{U}}(n) = \hat{\boldsymbol{q}}_{k, \text{U}}(n|n-1)}.
$}
\end{equation}

Finally, by substituting the updated state information into the mapping expressions from states to the target parameters, the updated Doppler shift, elevation angle, and azimuth angle can be obtained as
\begin{equation}\label{equa21}
\begin{cases}
\hat{{u}}_k(n) &= g_u(\hat{\boldsymbol{q}}_\text{S}(n), \hat{\boldsymbol{q}}_{k, \text{U}}(n)) \\
\hat{{\theta}}_{k, \text{E}}(n) &= g_\text{E}(\hat{\boldsymbol{q}}_\text{S}(n), \hat{\boldsymbol{q}}_{k, \text{U}}(n)) \\
\hat{{\theta}}_{k, \text{A}}(n) &= g_\text{A}(\hat{\boldsymbol{q}}_\text{S}(n), \hat{\boldsymbol{q}}_{k, \text{U}}(n))
\end{cases},
\end{equation}
and thus, the target parameter vector can be given as $\hat{\boldsymbol{z}}_k(n) = [\hat{{u}}_k(n), \hat{{\theta}}_{k, \text{E}}(n), \hat{{\theta}}_{k, \text{A}}(n)]^T$.

\subsection{CSI Acquisition}
According to the channel model in (\ref{equa27}), with the updated Doppler shift $\hat{u}_k(n)$ after the tracking, the CSI in the $n$-th block can be easily acquired by the LS method as
\begin{equation}\label{equa40}
\boldsymbol{\hat{h}}_k(n)= \frac{1}{N_\text{P}}\sum_{b=0}^{N_\text{P}-1}(\sqrt{P_k} \boldsymbol{W}_k)^\dagger e^{-j2\pi \hat{u}_k(n) b T_\text{sym}} {\boldsymbol{r}}_k(n,b).
\end{equation}

The procedure of the proposed JPCT algorithm for the $k$-th GU and in the $n$-th block is summarized in {\bf Algorithm 1}.

\begin{remark}
According to the steps of the proposed JPCT algorithm, it is obvious that the accuracy of the updated target parameters is directly related to the results of the EKF-based tracking. Therefore, the performance of the proposed algorithm depends on the state evolution noise and the accuracy of the predicted measurement error covariance, which will be investigated in the simulation part. Besides, compared with the rough estimation methods, the proposed algorithm with the same pilot overhead has a better performance based on the state tracking, while its overall complexity orders do not increase (see subsection \ref{Sec4}. F for details). Hence, the proposed algorithm has good performance with low pilot overhead and is quite applicable to the resource-limited LEO SATCOM systems.
\end{remark}

\begin{table*}[ht]
\caption{COMPLEXITY ANALYSIS}
\begin{center}
\begin{tabular}{|c|c|}
\hline
Algorithm & Overall complexity \\
\hline\hline
(1): Rough Doppler estimation & $\mathcal{O}(M_\text{RF}(N_\text{P}-1)^2+(N_\text{P}-1)^3 )$ \\
\hline
(2): Rough angle estimation & $\mathcal{O}(GMM_\text{RF}+GM_\text{RF}N_\text{P})$ \\
\hline
(3): Measurement error prediction & $\mathcal{O}(MM_\text{RF}^2+M_\text{RF}N_\text{P})$ \\
\hline
(4): State tracking and parameter update & $\mathcal{O}(1)$ \\
\hline
(5): CSI acquisition & $\mathcal{O}(MM_\text{RF}N_\text{P})$ \\
\hline
(1)+(2)+(5): Benchmark scheme & $\mathcal{O}(M_\text{RF}(N_\text{P}-1)^2+(N_\text{P}-1)^3 + GMM_\text{RF}+GM_\text{RF}N_\text{P})$\\
\hline
(1)+(2)+(3)+(4)+(5): Proposed JPCT algorithm & $\mathcal{O}(M_\text{RF}(N_\text{P}-1)^2 + (N_\text{P}-1)^3 + GMM_\text{RF} + GM_\text{RF}N_\text{P})$\\
\hline
\end{tabular}
\label{Complexity_Comparison}
\end{center}
\end{table*}

\begin{table}[ht]
\caption{SIMULATION PARAMETERS \cite{ALiao7,SHan9,BZheng11,MYing13}}
\begin{center}
\begin{tabular}{|c|c|}
\hline
Parameter & Value \\
\hline\hline
Carrier frequency $f_c$ & 1910 MHz \\
\hline
Carrier bandwidth $B_w$ & 4 MHz \\
\hline
Number of GUs $K$ & 1 \\
\hline
Number of propagation paths $L_k$ & 2 \\
\hline
Number of subcarriers $N_\text{sc}$ & 64 \\
\hline
\begin{tabular}{c}Number of antennas at LEO \\ satellite $M$ \end{tabular}& 64 \\
\hline
Number of RF chains $M_\text{RF}$ & 32 \\
\hline
Number of blocks $N_\text{B}$ & 10 \\
\hline
\begin{tabular}{c}Duration of one OFDM \\ symbol $T_\text{sym}$ \end{tabular}& $\frac{N_\text{sc}}{2B_w} = 8$ $\mu\text{s}$ \\
\hline
\begin{tabular}{c}Number of OFDM symbols \\ in one block $N_\text{ofdm}$ \end{tabular} & 312500 \\
\hline
Duration of one block $T_\text{B}$ & $N_\text{ofdm}T_\text{sym}= $ 2.5 s\\
\hline
Light speed $c$ & $3\times 10^8$ m/s \\
\hline
Transmit power $P_k$ & 30 dBm \\
\hline
Satellite antenna gain $\gamma_k$ & 8 dBi \\
\hline
Rician factor $\lambda_k$ & 8 \\
\hline
\begin{tabular}{c}
Ratio of GU antenna gain to \\ noise temperature $G_k/T$
 \end{tabular} & 1 dB/K \\
 \hline
Boltzmann's constant $\kappa$ & $1.38 \times 10^{-23}$ J/m \\
\hline
Height of satellite orbit $D$ & 600 km \\
\hline
Radius of earth $r_\text{E}$ & 6370 km \\
\hline
Velocity of LEO satellite $|v_\text{S}|$ & $7.6$ km/s \\
\hline
Velocity of GU $|v_{k,\text{U}}|$ & 100 km/h \\
\hline
\begin{tabular}{c}
Standard deviation of position \\ evolution noise for GU $\sigma_\text{U}$ \end{tabular}
& 10 m \\
\hline
\begin{tabular}{c}Standard deviation of velocity \\ evolution noise for GU $\sigma_v$ \end{tabular}
& 1 m/s \\
\hline
Number of pilots $N_\text{P}$ & 10 \\
\hline
\begin{tabular}{c}Resolution for rough angle \\ estimation $N_\text{E} = N_\text{A}$ \end{tabular} & 100 \\
\hline
\end{tabular}
\label{tab1}
\end{center}
\end{table}

\subsection{Complexity Analysis}
In this subsection, we will discuss the computational complexity of the proposed JPCT algorithm in the $n$-th block and for the $k$-th GU. As mentioned earlier, the proposed algorithm can be divided into five parts during each block: (1) rough Doppler estimation; (2) rough angle estimation; (3) measurement error prediction based on CRLB; (4) state tracking and parameter update; and (5) CSI acquisition. In the rough Doppler estimation, the main calculation lies in the steps of acquiring correlation matrices and (generalized) eigenvalue decomposition, which need about $\mathcal{O}(M_\text{RF}(N_\text{P}-1)^2)$ and $\mathcal{O}((N_\text{P}-1)^3)$ calculations \cite{YH28}, and thereby its total calculations are $\mathcal{O}(M_\text{RF}(N_\text{P}-1)^2+(N_\text{P}-1)^3 )$. In the rough angle estimation, the SOMP algorithm requires the complexity of $\mathcal{O}(GMM_\text{RF}+GM_\text{RF}N_\text{P})$. As to the measurement error prediction, the prediction of summation term requires the calculations of $\mathcal{O}(MM_\text{RF}+M_\text{RF}N_\text{P})$, and the complexity of CRLB calculations is $\mathcal{O}(MM_\text{RF}^2)$. Thus, the total complexity of the measurement error prediction is $\mathcal{O}(MM_\text{RF}^2+M_\text{RF}N_\text{P})$. As to the EKF-based state tracking, due to the limited dimension of the tracked states, the complexity of the state tracking and parameter update is in a constant order, which can be neglected when other parameters are large enough. In the CSI acquisition, the LS process needs about $\mathcal{O}(MM_\text{RF}N_\text{P})$ calculations with the modified Gram-Schmidt algorithm and QR factorization employed \cite{JLee32}. In conclusion, the overall computational complexity of the proposed algorithm is about $\mathcal{O}(M_\text{RF}(N_\text{P}-1)^2 + (N_\text{P}-1)^3 + GMM_\text{RF} + GM_\text{RF}N_\text{P})$ with the assumption of $G>M>M_\text{RF}$.

In order to evaluate the complexity performance of the proposed algorithm, we employ a benchmark scheme with the rough estimation methods. Specifically, the Doppler shift and AoAs can be obtained by the ESPRIT and SOMP algorithms, respectively. Subsequently, with the rough Doppler shift acquired, the channel can be estimated by the LS technique as (\ref{equa40}) similarly. Therefore, the benchmark scheme for the parameter and channel estimation can be viewed as a combination of parts (1), (2), and (5) of the proposed JPCT algorithm. For convenience, a table for complexity analysis is given in TABLE \ref{Complexity_Comparison}. From TABLE \ref{Complexity_Comparison}, it can be easily observed that compared to the benchmark scheme, the increase of complexity for the proposed algorithm mainly lies in the measurement error prediction with complexity of $\mathcal{O}(MM_\text{RF}^2+M_\text{RF}N_\text{P})$, which will not increase the overall complexity and thus is acceptable in the resource-limited LEO SATCOM systems.

\section{Simulation Results}\label{Sec5}
%\vspace{-0.5cm}
\begin{figure*}[t]
	\centering
	\vspace{-0.15in}
	\begin{minipage}{1\linewidth}
        \subfigure[]{
			\label{Fig3_Doppler} \includegraphics[width=0.34\linewidth]{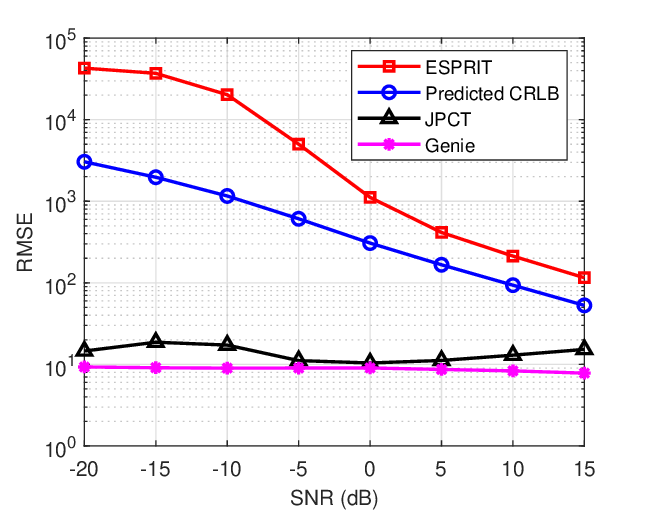} }\hspace{-0.95em}%
		\subfigure[]{
			\label{Fig3_Elevation}	\includegraphics[width=0.34\linewidth]{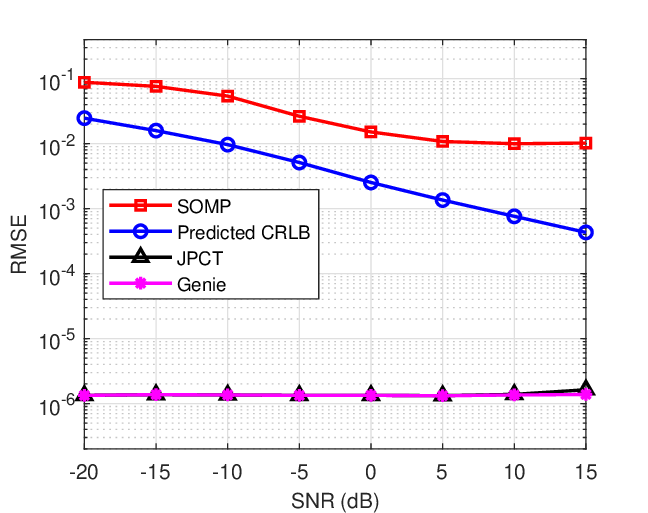}
		}\hspace{-0.95em}%
		\subfigure[]{
			\label{Fig3_Azimuth}		\includegraphics[width=0.34\linewidth]{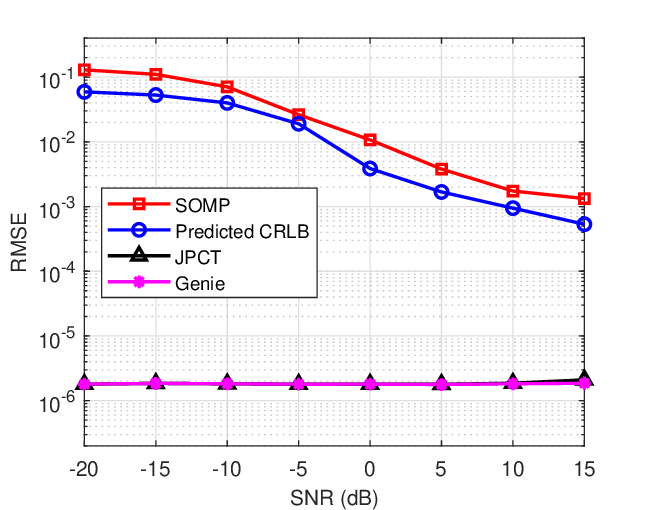}
		}
	\end{minipage}
	\caption{RMSE performance against SNRs for: (a) Doppler shift; (b) elevation angle; and (c) azimuth angle.}
	%\vspace{-0.2in}
	\label{Fig_3}
\end{figure*}

\begin{figure*}[t]
	\centering
	\vspace{-0.15in}
	\begin{minipage}{1\linewidth}
		\subfigure[]{
			\label{Fig4_Doppler} \includegraphics[width=0.34\linewidth]{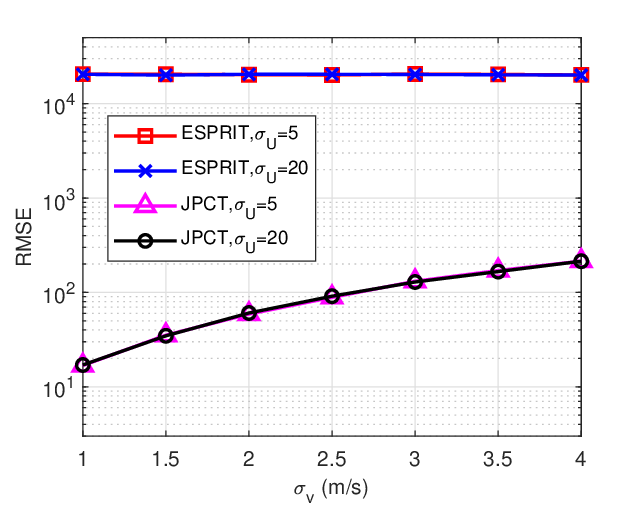} }\hspace{-0.95em}%
		\subfigure[]{
			\label{Fig4_Elevation}	\includegraphics[width=0.34\linewidth]{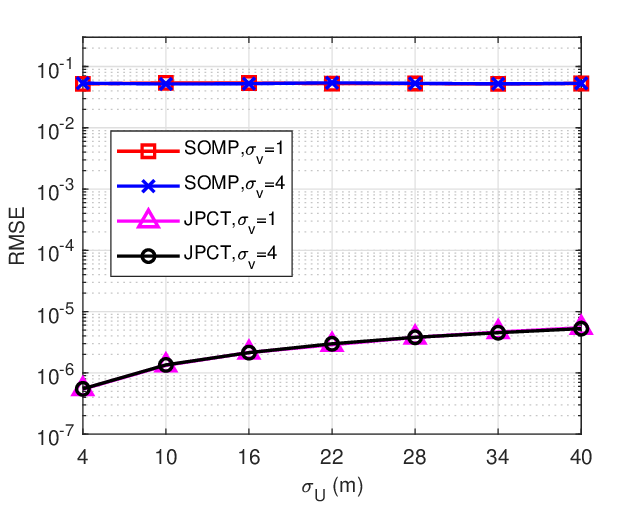}
		}\hspace{-0.95em}%
		\subfigure[]{
			\label{Fig4_Azimuth}		\includegraphics[width=0.34\linewidth]{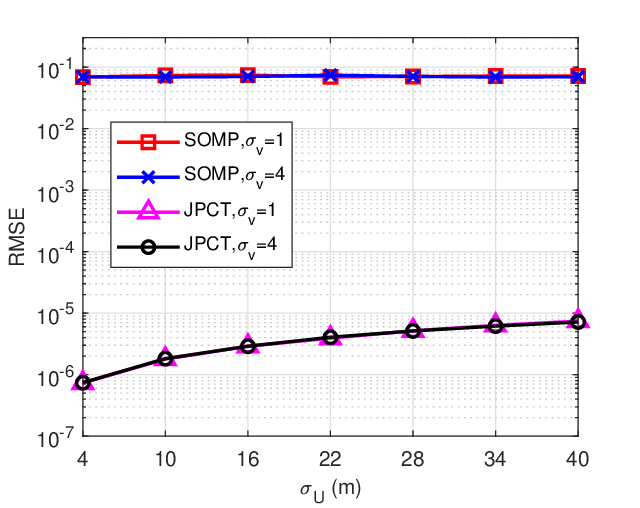}
		}
	\end{minipage}
	\caption{Analysis on effects of state evolution noise with $\text{SNR} = -10$ dB: (a) RMSE performance against $\sigma_v$ with $\sigma_\text{U}\in\{5,20\}$ for Doppler shift; RMSE performance against $\sigma_\text{U}$ with $\sigma_v \in \{1, 4\}$ for: (b) elevation angle; and (c) azimuth angle.}
	%\vspace{-0.2in}
	\label{Fig_4}
\end{figure*}

In this section, simulation results are presented to evaluate the performance of the proposed JPCT algorithm. We consider an LEO SATCOM system under the direct communication scenario, where the carrier frequency and bandwidth are set as $f_c=1910$ MHz and $B_w=4$ MHz, according to the personal communications service (PCS) band adopted in Starlink's direct communication project \cite{RBrome34}. Moreover, a UPA with $M=M_x\times M_y = 8\times 8 = 64$ antennas is mounted on the satellite and the number of RF chains is $M_\text{RF}=32$. Note that with the OFDM system with $N_\text{sc} = 64$ subcarriers, we only consider one GU that transmits the signal over the $2$-nd subcarrier in simulation ($m=1$), while the multi-user scenario is also applicable. The orbit height of the LEO satellite is set as $D = 600$ km with the earth radius $r_\text{E} = 6370$ km. Without loss of generality, we assume that the satellite orbit passes through the X-axis with the inclination angle of $53^{\circ}$, and the initial position of the satellite is set as $\boldsymbol{p}_\text{S}(0) = [(r_\text{E}+D),0,0]^T$. With the initial position of GU as $\boldsymbol{p}_\text{U}(0) = [x_u,y_u,z_u]^T = 10^6 \times[5,2.7908,2.7908]^T$, we can easily set a moving orbit for GU with its normal vector of $[1,-1,\frac{(y_u-x_u)}{z_u}]^T$. Furthermore, the UPA plane is assumed to be vertical to the satellite orbit plane and always directs to the center of the earth. As to the parametric channel model, the state-dependent parameters including AoAs and Doppler shift can be determined by the states of the satellite and GUs, while the channel gains and delays can be generated by random Gaussian distribution and uniform distribution, respectively \cite{ALiao7}. For simplicity, we assume that the variances of the state evolution noise of GUs for different directions are the same, namely $\sigma_{x_\text{u}}^2 = \sigma_{y_\text{u}}^2 = \sigma_{z_\text{u}}^2 = \sigma_\text{U}^2$ and $\sigma_{v_x}^2 = \sigma_{v_y}^2 = \sigma_{v_z}^2 = \sigma_v^2$. Unless specified otherwise, the detailed parameter settings in the simulation are summarized in TABLE \ref{tab1}. Besides, we adopt the root mean square error (RMSE) and normalized mean square error (NMSE) as the performance metrics for parameter and channel tracking respectively \cite{ALiao7, FLiu16}, which can be described as
\begin{equation}\label{equa23}
  \text{RMSE}_x = \frac{1}{N_\text{B}} \sum_{n=0}^{N_\text{B}-1}\sqrt{\mathbb{E}\{|x_n-\hat{x}_n|^2\}},
\end{equation}
\begin{equation}
\text{NMSE} = \frac{1}{N_\text{B}} \sum_{n=0}^{N_\text{B}-1}\frac{\|\boldsymbol{h}(n) - \hat{\boldsymbol{h}}(n)\|^2}{\|\boldsymbol{h}(n)\|^2},
\end{equation}
where $x_n = \{u(n), \theta_\text{E}(n), \theta_\text{A}(n)\}$. $\boldsymbol{h}(n)$ and $\hat{\boldsymbol{h}}(n)$ denote the real and estimated channel, respectively.

Fig. \ref{Fig_3} depicts the RMSE curves of the target parameters against different SNRs ranging from -20 dB to 15 dB. Notice that ``ESPRIT'' and ``SOMP'' represent the rough estimation methods for Doppler shift and channel angles, respectively. In each subfigure, as to the corresponding estimation methods for Doppler shift, elevation angle, and azimuth angle, ``Predicted CRLB'', ``JPCT'', and ``Genie'' denote the predicted CRLB with rough estimates, the proposed algorithm, and the proposed algorithm with the ideal case that the measurement error covariance can be predicted perfectly, respectively. It can be observed that the predicted CRLB can describe the lower bound of the rough estimation methods, which validates the correctness of the formulation. In both the estimation of Doppler shift and channel angles, the proposed JPCT algorithm can highly improve the RMSE performance compared with the rough estimation methods, even with the same pilot overhead. However, the proposed algorithm may suffer a little performance degradation when the mismatch of the predicted measurement error covariance is relatively large, which can be shown from the gap between curves of the rough estimation methods and predicted CRLB in each subfigure. It can also be verified from the curves of ``Genie''. With the perfect measurement error prediction, the ideal case presents good performance in all SNR regions. Besides, the proposed algorithm can effectively improve the RMSE performance, especially at low SNRs. This is due to the fact that a poor channel environment in low-SNR regions leads to the performance degradation of rough estimation methods, while the proposed JPCT algorithm almost maintains good performance in all SNR regions with accurate measurement error prediction. Therefore, the proposed algorithm can well adapt to the practical LEO SATCOM systems, which have large path losses and generally work in low-SNR environments.

Fig. \ref{Fig_4} analyzes the effects of state evolution noise (including the standard deviations of position evolution noise $\sigma_\text{U}$ and velocity evolution noise $\sigma_v$) on the RMSE performance, with the setting of $\text{SNR} = -10$ dB.  Fig. \ref{Fig4_Doppler} shows the RMSE curves for Doppler tracking versus  $\sigma_v$ with $\sigma_\text{U}\in \{5,20\}$. As to the tracking performance of AoAs, Fig. \ref{Fig4_Elevation} and Fig. \ref{Fig4_Azimuth} show the RMSE curves versus $\sigma_\text{U}$ with $\sigma_v \in \{1, 4\}$ for elevation angle and azimuth angle, respectively. From Fig. \ref{Fig_4}, it is clear that similar to the conclusions in Fig. \ref{Fig_3}, the proposed JPCT algorithm performs much better than the rough estimation methods even with large state evolution noise, which further validates the effectiveness of the proposed algorithm. In Fig. \ref{Fig4_Doppler}, although the Doppler shift depends on both the velocity and position of the GU, it can be shown that only the velocity evolution noise $\sigma_v$ affects the performance of Doppler tracking, where the RMSE grows with larger $\sigma_v$ while is independent of $\sigma_\text{U}$. In Fig. \ref{Fig4_Elevation} and Fig. \ref{Fig4_Azimuth}, since the expressions of AoAs are only related to the position parameters of the GU, it is obvious that the RMSE performance only depends on the position evolution noise of GU $\sigma_\text{U}$, where the tracking performance degrades with larger $\sigma_\text{U}$ while is independent of $\sigma_v$.

\begin{figure}[t]
  \centering
  \includegraphics[width=0.45\textwidth]{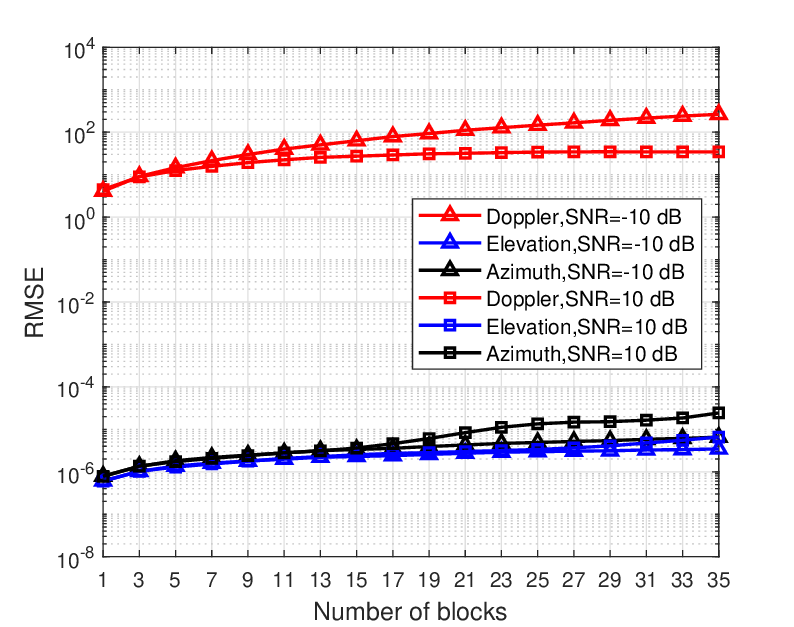}
  \caption{RMSE performance against number of blocks for target parameters with SNR$\in\{-10, 10\}$ dB.}\label{NB_RMSE}
\end{figure}

To investigate the performance of the proposed JPCT algorithm during each block, we update the RMSE metric in the $n$-th block as $\text{RMSE}_x(n) = \sqrt{\mathbb{E}\{|x_n-\hat{x}_n|^2\}}$, and depicts the RMSE curves for the target parameters including Doppler shift, elevation angle, and azimuth angle in Fig. \ref{NB_RMSE}, with SNR$\in \{-10, 10\}$ dB.  From Fig. \ref{NB_RMSE}, it is obvious that as the number of blocks grows, the RMSE of these three parameters all has a slight increment. This is due to the fact that the performance of the proposed algorithm greatly depends on the results of the EKF-based tracking. When blocks increase, the tracking error will be accumulated, which further leads to the performance degradation of the proposed algorithm. Moreover, the RMSE curves of AoAs with SNR=-10 dB have slighter performance degradation than curves with SNR=10 dB, while the RMSE curves for Doppler shift perform better at SNR=10 dB than that at SNR=-10 dB. This is due to the fact that the proposed algorithm is largely affected by the accuracy of the predicted measurement error covariance, resulting in a better performance at SNR=10 dB for Doppler shift while a better performance at SNR=-10 dB for AoAs, according to the results in Fig. \ref{Fig_3}.

\begin{figure}[t]
  \centering
  \includegraphics[width=0.45\textwidth]{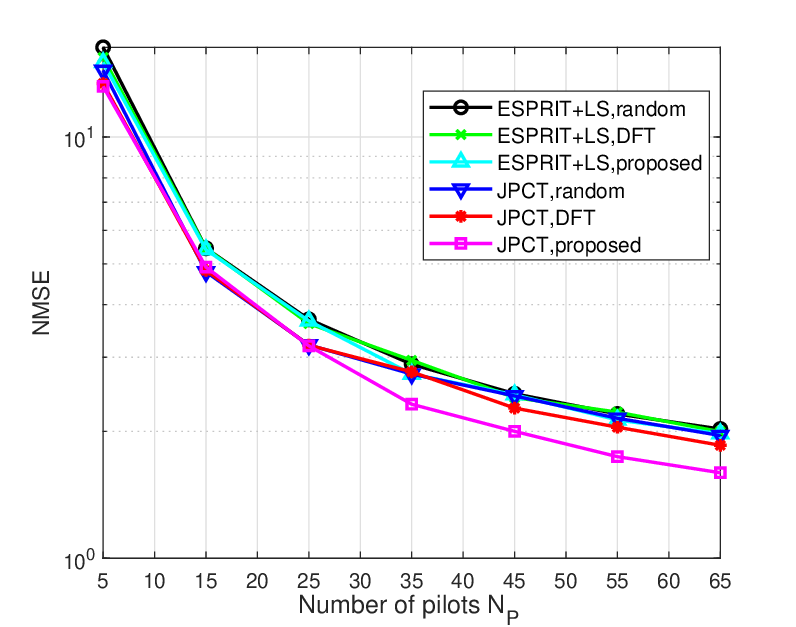}
  \caption{NMSE performance against number of pilots $N_\text{P}$ for CSI acquisition with different combiners and SNR=-10 dB.}\label{Pilot_RMSE}
\end{figure}

As to the performance of the CSI acquisition under the proposed JPCT algorithm, Fig. \ref{Pilot_RMSE} studies the effects of the number of pilots $N_\text{P}$ on the NMSE performance of channel tracking with SNR=-10 dB. We consider a benchmark algorithm ``ESPRIT+LS'', where the Doppler shift is estimated by the ESPRIT algorithm and the CSI can be acquired by the LS algorithm as (\ref{equa40}) similarly. Meanwhile, different design methods for combiners are also investigated, where ``random'' and ``DFT'' denote the random and DFT matrices designed as the combiners, and ``proposed'' denotes the proposed combiner matrix as (\ref{equa32}). In Fig. \ref{Pilot_RMSE}, it can be observed that as the number of pilots $N_\text{P}$ increases, the NMSE performance for these algorithms all gradually improves. For the benchmark algorithm ``ESPRIT+LS'', the employment of random and DFT combiners indicates similar performance, while the proposed combiner leads to a slightly better performance. For the proposed algorithm, curves of the JPCT algorithm with these three combiners are all superior to the benchmark algorithms. Furthermore, the JPCT algorithm with the proposed combiner has the optimum performance, followed by the DFT and random combiners. In conclusion, under a given requirement for NMSE performance, the JPCT algorithm with the proposed combiner can reduce the pilot overhead compared to the conventional schemes without the EKF-based parameter tracking, which is of great benefit in the resource-limited SATCOM systems.

\section{Conclusion}\label{Sec6}
In this paper, we proposed a joint parameter and channel tracking algorithm for LEO SATCOM systems by exploiting the on-orbit characteristics of LEO satellite and GUs. Simulation results showed that the proposed algorithm highly outperformed the corresponding rough estimation methods, even with the same pilot overhead. Besides, it also validated that the proposed algorithm can be well applied to the LEO SATCOM systems under low-SNR environments. Besides, more advanced measurement error prediction methods and more practical scenarios with multi-user interference can be investigated in future works.

\nocite{*}
\bibliographystyle{IEEE}

\begin{thebibliography}{40}

\bibitem{JChu35} % 1
J. Chu, X. Chen, C. Zhong, and Z. Zhang, ``Robust design for NOMA-based multibeam LEO satellite Internet of things,'' \emph{IEEE Internet Things J.}, vol. 8, no. 3, pp. 1959-1970, Feb. 2021.

\bibitem{YHe40} % 2
Y. He, Y. Liu, C. Jiang, and X. Zhong, ``Multiobjective anti-collision for massive access ranging in MF-TDMA satellite communication system,'' \emph{IEEE Internet Things J.}, vol. 9, no. 16, pp. 14655-14666, Aug. 2022.

\bibitem{ZQu41} % 3
Z. Qu, G. Zhang, H. Cao, and J. Xie, ``LEO satellite constellation for Internet of Things,'' \emph{IEEE Access}, vol. 5, pp. 18391-18401, Aug. 2017.

\bibitem{JLiu2} % 4
J. Liu, Y. Shi, Z. M. Fadlullah, and N. Kato, ``Space-air-ground integrated network: A survey,'' \emph{IEEE Commun. Survs. Tuts.}, vol. 20, no. 4, pp. 2714-2741, 4th Quart., 2018.

\bibitem{ZLin42} % 5
Z. Lin, M. Lin, B. Champagne, W.-P. Zhu, and N. Al-Dhahir, ``Secrecy-energy efficient hybrid beamforming for satellite-terrestrial integrated networks,'' \emph{IEEE Trans. Commun.}, vol. 69, no. 9, pp. 6345-6360, Sep. 2021.

\bibitem{LYou3} % 6
L. You, X. Qiang, K. -X. Li, C. G. Tsinos, W. Wang, X. Gao, and B. Ottersten, ``Hybrid analog/digital precoding for downlink massive MIMO LEO satellite communications,'' \emph{IEEE Trans. Wireless Commun.}, vol. 21, no. 8, pp. 5962-5976, Aug. 2022.

\bibitem{HAl4} % 7
H. Al-Hraishawi, H. Chougrani, S. Kisseleff, E. Lagunas, and S. Chatzinotas, ``A survey on nongeostationary satellite systems: The communication perspective,'' \emph{IEEE Commun. Survs. Tuts.}, vol. 25, no. 1, pp. 101-132, 1st Quart., 2023.

\bibitem{Bwang33} % 8
B. Wang, SpaceX and T-Mobile Will Put Cellphone Coverage into Starlink Gen 2, Nextbigfuture, Aug. 25, 2022. [Online]. Available: https://www.nextbigfuture.com/2022/08/spacex-and-t-mobile-will-put-cellphone-coverage-into-starlink-gen-2.html.

\bibitem{JHeo5} % 9
J. Heo, S. Sung, H. Lee, I. Hwang, and D. Hong, ``MIMO satellite communication systems: A survey from the PHY layer perspective,'' \emph{IEEE Commun. Survs. Tuts.}, vol. 25, no. 3, pp. 1543-1570, 3rd Quart., 2023.

\bibitem{SSri17} % 10
S. Srivastava, C. S. K. Patro, A. K. Jagannatham, and L. Hanzo, ``Sparse, group-sparse, and online Bayesian learning aided channel estimation for doubly-selective mmWave hybrid MIMO OFDM systems,'' \emph{IEEE Trans. Commun.}, vol. 69, no. 9, pp. 5843-5858, Sep. 2021.

\bibitem{JRF18} % 11
J. Rodríguez-Fernández, N. González-Prelcic, K. Venugopal, and R. W. Heath, ``Frequency-domain compressive channel estimation for frequency-selective hybrid millimeter wave MIMO systems,'' \emph{IEEE Trans. Wireless Commun.}, vol. 17, no. 5, pp. 2946-2960, May 2018.

\bibitem{ALiao19} % 12
A. Liao, Z. Gao, H. Wang, S. Chen, M. -S. Alouini, and H. Yin, ``Closed-loop sparse channel estimation for wideband millimeter-wave full-dimensional MIMO systems,'' \emph{IEEE Trans. Commun.}, vol. 67, no. 12, pp. 8329-8345, Dec. 2019.

\bibitem{ZGuo20} % 13
Z. Guo, X. Wang, and W. Heng, ``Millimeter-wave channel estimation based on 2-D beamspace MUSIC method,'' \emph{IEEE Trans. Wireless Commun.}, vol. 16, no. 8, pp. 5384-5394, Aug. 2017.

\bibitem{RZhang21} % 14
R. Zhang, L. Cheng, S. Wang, Y. Lou, W. Wu, and D. W. K. Ng, ``Tensor decomposition-based channel estimation for hybrid mmWave massive MIMO in high-mobility scenarios,'' \emph{IEEE Trans. Commun.}, vol. 70, no. 9, pp. 6325-6340, Sep. 2022.

\bibitem{YLin22} % 15
Y. Lin, S. Jin, M. Matthaiou, and X. You, ``Tensor-based channel estimation for millimeter wave MIMO-OFDM with dual-wideband effects,'' \emph{IEEE Trans. Commun.}, vol. 68, no. 7, pp. 4218-4232, Jul. 2020.

\bibitem{AAbda23} % 16
A. Abdallah, A. Celik, M. M. Mansour, and A. M. Eltawil, ``Deep learning-based frequency-selective channel estimation for hybrid mmWave MIMO systems,'' \emph{IEEE Trans. Wireless Commun.}, vol. 21, no. 6, pp. 3804-3821, Jun. 2022.

\bibitem{XMa24} % 17
X. Ma, Z. Gao, F. Gao, and M. Di Renzo, ``Model-driven deep learning based channel estimation and feedback for millimeter-wave massive hybrid MIMO systems,'' \emph{IEEE J. Sel. Areas Commun.}, vol. 39, no. 8, pp. 2388-2406, Aug. 2021.

\bibitem{JGao36} % 18
J. Gao, C. Zhong, G. Y. Li, J. B. Soriaga, and A. Behboodi, ``Deep learning-based channel estimation for wideband hybrid mmWave massive MIMO,'' \emph{IEEE Trans. Commun.}, vol. 71, no. 6, pp. 3679-3693, Jun. 2023.

\bibitem{LYou6} % 19
L. You, K. -X. Li, J. Wang, X. Gao, X. -G. Xia, and B. Ottersten, ``Massive MIMO transmission for LEO satellite communications,'' \emph{IEEE J. Sel. Areas Commun.}, vol. 38, no. 8, pp. 1851-1865, Aug. 2020.

\bibitem{YZhang25} % 20
Y. Zhang, Y. Wu, A. Liu, X. Xia, T. Pan, and X. Liu, ``Deep learning-based channel prediction for LEO satellite massive MIMO communication system,'' \emph{IEEE Wireless Commun. Lett.}, vol. 10, no. 8, pp. 1835-1839, Aug. 2021.

\bibitem{ALiao7} % 21
A. Liao, Z. Gao, D. Wang, H. Wang, H. Yin, D. W. K. Ng, and M. -S. Alouini, ``Terahertz ultra-massive MIMO-based aeronautical communications in space-air-ground integrated networks,'' \emph{IEEE J. Sel. Areas Commun.}, vol. 39, no. 6, pp. 1741-1767, Jun. 2021.

\bibitem{YLiu8} % 22
Y. Liu, Y. Su, Y. Zhou, H. Cao, and J. Shi, ``Frequency offset estimation for high dynamic LEO satellite communication systems,'' in \emph{2019 11th International Conference on Wireless Communications and Signal Processing (WCSP)}, 2019, pp. 1-6.

\bibitem{SHan9} % 23
S. Han, W. Lee, W. Shin, and J. -H. Kim, ``Adaptive beam size design for LEO satellite networks with Doppler shift compensation,'' in \emph{2022 IEEE VTS Asia Pacific Wireless Communications Symposium (APWCS)}, 2022, pp. 26-30.

\bibitem{JYu10} % 24
J. Yu, X. Liu, Y. Gao, and X. Shen, ``3D channel tracking for UAV-satellite communications in space-air-ground integrated networks,'' \emph{IEEE J. Sel. Areas Commun.}, vol. 38, no. 12, pp. 2810-2823, Dec. 2020.

\bibitem{BZheng11} % 25
B. Zheng, S. Lin, and R. Zhang, ``Intelligent reflecting surface-aided LEO satellite communication: Cooperative passive beamforming and distributed channel estimation,'' \emph{IEEE J. Sel. Areas Commun.}, vol. 40, no. 10, pp. 3057-3070, Oct. 2022.

\bibitem{TYue12} % 26
T. Yue, A. Liu, and X. Liang, ``Block-based Kalman channel tracking for LEO satellite communication with massive MIMO,'' \emph{IEEE Commun. Lett.}, vol. 27, no. 2, pp. 645-649, Feb. 2023.

\bibitem{ZWei14} % 27
Z. Wei, F. Liu, C.Liu, Z. Yang, D. W. K. Ng, and R. Schober, ``Integrated sensing, navigation, and communication for secure UAV networks with
a mobile eavesdropper,'' 2023, \emph{arXiv:2305.12842}.

\bibitem{FLiu16}% 28
F. Liu, W. Yuan, C. Masouros, and J. Yuan, ``Radar-assisted predictive beamforming for vehicular links: Communication served by sensing,'' \emph{IEEE Trans. Wireless Commun.}, vol. 19, no. 11, pp. 7704-7719, Nov. 2020.

\bibitem{XHu39} % 29
X. Hu, F. Gao, C. Zhong, X. Chen, Y. Zhang, and Z. Zhang, ``An angle domain design framework for intelligent reflecting surface systems,'' in \emph{2020 IEEE Global Communications Conference (GLOBECOM 2020)}, Taipei, Taiwan, 2020, pp. 1-6.

\bibitem{MYing13}% 30
M. Ying, X. Chen, and X. Shao, ``Exploiting tensor-based Bayesian learning for massive grant-free random access in LEO satellite Internet of Things,'' \emph{IEEE Trans. Commun.}, vol. 71, no. 2, pp. 1141-1152, Feb. 2023.

\bibitem{MHuang29} % 31
M. Huang, J. Chen, and S. Feng, ``Synchronization for OFDM-based satellite communication system,'' \emph{IEEE Trans. Veh. Technol.}, vol. 70, no. 6, pp. 5693-5702, Jun. 2021.

\bibitem{IAli15} % 32
I. Ali, N. Al-Dhahir, and J. E. Hershey, ``Doppler characterization for LEO satellites,'' \emph{IEEE Trans. Commun.}, vol. 46, no. 3, pp. 309-313, Mar. 1998.

\bibitem{KMoon37} % 33
K. Moon, H. Kwon, C. -K. Ryoo, and H. Sim, ``Trajectory estimation for a ballistic missile in ballistic phase using IR images,'' in \emph{2018 9th International Conference on Mechanical and Aerospace Engineering (ICMAE)}, Budapest, Hungary, 2018, pp. 173-177.

\bibitem{XHY38} % 34
X. -H. Yan, B. -G. Cai, B. Ning, and W. ShangGuan, ``Moving horizon optimization of dynamic trajectory planning for high-speed train operation,'' \emph{IEEE Trans. Intell. Transp. Syst.}, vol. 17, no. 5, pp. 1258-1270, May 2016.

\bibitem{RRoy26} % 35
R. Roy, A. Paulraj, and T. Kailath, ``ESPRIT–A subspace rotation approach to estimation of parameters of cisoids in noise,'' \emph{IEEE Trans. Acoust., Speech, Signal Process.}, vol. 34, no. 5, pp. 1340–1342, Oct. 1986.

%\bibitem{RSch27}
%R. Schmidt, ``Multiple emitter location and signal parameter estimation,'' \emph{IEEE Trans. Antennas Propag.}, vol. 34, no. 3, pp. 276–280, Mar. 1986.

\bibitem{CLin30} % 36
C. Lin, J. Gao, R. Jin, and C. Zhong, ``Self-adaptive measurement matrix design and channel estimation in time-varying hybrid mmWave massive MIMO-OFDM systems,'' \emph{IEEE Trans. Commun.}, vol. 72, no. 1, pp. 618-629, Jan. 2024.

\bibitem{ZGao31} % 37
Z. Gao, L. Dai, Z. Wang, and S. Chen, ``Spatially common sparsity based adaptive channel estimation and feedback for FDD massive MIMO,'' \emph{IEEE Trans. Signal Process.}, vol. 63, no. 23, pp. 6169-6183, Dec., 2015.

\bibitem{YH28} % 38
Y. -H. Shao, N. -Y. Deng, W. -J. Chen, and Z. Wang, ``Improved generalized eigenvalue proximal support vector machine,'' \emph{IEEE Signal Process. Lett.}, vol. 20, no. 3, pp. 213-216, Mar. 2013.

\bibitem{JLee32} % 39
J. Lee, G. -T. Gil, and Y. H. Lee, ``Channel estimation via orthogonal matching pursuit for hybrid MIMO systems in millimeter wave communications,'' \emph{IEEE Trans. Commun.}, vol. 64, no. 6, pp. 2370-2386, Jun. 2016.

\bibitem{RBrome34} % 40
R. Brome, T-Mobile Teams with SpaceX for Satellite Coverage Using Existing Phones, Phonescoop, Aug. 26, 2022. [Online]. Available: https://www.phonescoop.com/articles/article.php?a=22860.

\end{thebibliography}
\begin{footnotesize}

\end{footnotesize}

\end{document}